\documentclass[12pt,preprint]{aastex}
\usepackage{multirow}

\newcommand{\degrees}{\ensuremath{^\circ}}
\shorttitle{Molecular gas in z$\sim$6 quasars}
\shortauthors{Wang et al.}

\begin{document}

\title{Molecular Gas in Redshift 6 Quasar Host Galaxies}

\author{Ran Wang\altaffilmark{1, 2},
Chris L. Carilli\altaffilmark{2},
R. Neri\altaffilmark{3},
D. A. Riechers\altaffilmark{4,11}
Jeff Wagg\altaffilmark{2,5},
Fabian Walter\altaffilmark{6},
Frank Bertoldi\altaffilmark{7},
Karl M. Menten\altaffilmark{8},
Alain Omont\altaffilmark{9},
Pierre Cox\altaffilmark{3},
Xiaohui Fan\altaffilmark{10}}
\altaffiltext{1}{Purple Mountain Observatory,China Academy of Science, No. 2 Beijing West Road, Nanjing, 210008, China}
\altaffiltext{2}{National Radio Astronomy Observatory, PO Box 0, Socorro, NM, USA 87801}
\altaffiltext{3}{Institute de Radioastronomie Millimetrique, St. Martin d'Heres, F-38406, France}
\altaffiltext{4}{California Institute of Technology, 1200 E. California Blvd., Pasadena, CA, 91125, USA}
\altaffiltext{5}{European Southern Observatory, Alonso de C\'ordova 3107, Vitacura, Casilla 19001, Santiago 19, Chile}
\altaffiltext{6}{Max-Planck-Institute for Astronomy, K$\rm \ddot o$nigsstuhl 17, 69117 Heidelberg, Germany}
\altaffiltext{7}{Argelander-Institut f$\rm \ddot u$r Astronomie, University of Bonn, Auf dem H$\rm \ddot u$gel 71, 53121 Bonn, Germany}
\altaffiltext{8}{Max-Planck-Institut f$\rm \ddot u$r Radioastronomie, Auf dem H$\rm \ddot u$gel 71, 53121 Bonn, Germany}
\altaffiltext{9}{Institut d'Astrophysique de Paris, CNRS and Universite Pierre et Marie Curie, Paris, France}
\altaffiltext{10}{Steward Observatory, The University of Arizona, Tucson, AZ 85721}
\altaffiltext{11}{Hubble Fellow}
\begin{abstract}
We report our new observations of redshifted carbon monoxide emission 
from six z$\sim$6 quasars, using the IRAM Plateau de Bure Interferometer. 
CO (6-5) or (5-4) line emission was detected in all six sources. 
Together with two other 
previous CO detections, these observations provide unique constraints 
on the molecular gas emission properties in these quasar systems 
close to the end of the cosmic reionization. Complementary results 
are also presented for low-J CO lines observed at the Green Bank 
Telescope and the Very Large Array, and dust continuum 
from five of these sources with the SHARC-II bolometer camera 
at the Caltech Submillimeter Observatory. We then present 
a study of the molecular gas properties in our combined 
sample of eight CO-detected quasars at z$\sim$6. 
The detections of high-order CO line emission in these 
objects indicates the presence of highly excited molecular gas, with estimated masses 
on the order of $\rm 10^{10}\,M_{\odot}$ within the quasar host galaxies. 
No significant difference is found in the gas mass and CO 
line width distributions between our $z \sim 6$ quasars and samples of 
CO-detected $\rm 1.4\leq z\leq5$ quasars and submillimeter galaxies. 
Most of the CO-detected quasars at z$\sim$6 
follow the far infrared-CO luminosity relationship defined by actively star-forming 
galaxies at low and high redshifts. This suggests that ongoing star formation 
in their hosts contributes significantly to 
the dust heating at FIR wavelengths. The result is consistent with the picture 
of galaxy formation co-eval with supermassive black hole (SMBH) 
accretion in the earliest quasar-host systems. 
We investigate the black hole--bulge 
relationships of our quasar sample, 
using the CO dynamics as a tracer for the dynamical mass of the quasar host. 
The median estimated black hole-bulge mass ratio is about fifteen times higher 
than the present-day value of $\sim$0.0014. This places important  
constraints on the formation and evolution of the 
most massive SMBH-spheroidal host systems at the highest redshift. 
\end{abstract}

\section{Introduction}

Optically bright and massive quasar systems have been discovered 
at z$\sim$6 through large optical surveys (e.g., Fan et al. 
2006a; Willott et al. 2007, 2009a, 2009b; Cool et al. 2006; 
Jiang et al. 2008). The Gunn-Peterson absorption seen in their 
rest-frame UV spectra indicates that they exist close to the 
epoch of the cosmic reionization, and thus, are among the first 
generation of the most luminous objects in the universe (Fan et al. 2006b; 
Fan et al. 2006c). About 30\% of the z$\sim$6
quasars have been detected in dust continuum emission at 250 GHz, 
indicating large far-infrared (FIR) luminosities ($\rm 3\times10^{12}\,L_{\odot}$ 
to $\rm 10\times10^{12}\,L_{\odot}$, Wang et al. 2008b)
 arising from dust with temperatures of 30 to 60 K (e.g., 
Benford et al. 1999; Petric et al. 2003; Bertoldi et al. 2003a;
Beelen et al. 2006; Wang et al. 2007, 2008b). Studies of these 
objects' optical-to-radio spectral energy distributions (SEDs) 
and luminosity correlations argue for dust heating by star 
formation in the host galaxies (Bertoldi et al. 2003a; 
Beelen et al. 2006; Carilli et al. 2004; Riechers et al. 2007; 
Wang et al. 2008b). High resolution imaging of the dust continuum 
and [C II] $\rm \lambda\,158\mu m$ line emission in the brightest 
z$\sim$6 millimeter source, the z=6.42 quasar SDSS J114816.64+525150.3 
(hereafter J1148+5251), suggest a high star formation surface 
density in the quasar host of 
$\rm \sim1000\,M_{\odot}\,yr^{-1}\,kpc^{-2}$ 
(Walter et al. 2009). It is likely that we are witnessing 
an early galaxy evolutionary stage in these FIR luminous 
quasar systems at z$\sim$6, in which a massive starburst 
is ongoing co-evally with a luminous central active galactic 
nuclei (AGN). These objects therefore may also provide 
crucial insight into our understanding of the origin of the 
correlation between bulge mass/luminosity/stellar velocity 
dispersion and supermassive black hole (SMBH) mass in nearby
galaxies (Tremaine et al. 2002; Marconi \& Hunt 2003;
Hopkins et al. 2007a), which suggests that the
formation of the SMBH and its spheroidal host are tightly
coupled (Kauffmann \& Haehnelt 2000; Hopkins et al. 2007b).

Molecular CO line emission has been widely detected and studied 
in similar FIR luminous quasar systems at z$\rm >1$ to 5 (e.g., 
Brown \& Vanden Bout 1992; Omont et al. 1996a; 
Carilli et al. 2002; Cox et al. 2002; Solomon \& 
Vanden Bout 2005, hereafter SV05; Riechers et al. 2006; 
Maiolino et al. 2007; Coppin et al. 2008a). 
These CO detected quasars show a FIR-to-CO luminosity 
correlation similar to local starburst spiral galaxies, 
Ultra Luminous Infra Red Galaxies (ULIRGs), and 
high-z submillimeter galaxies (SMGs, SV05; 
Greve et al. 2005; Riechers et al. 2006). The derived  
molecular gas masses are on the order of $\rm \sim10^{10}\,M_{\odot}$, 
which are also comparable to the typical values found in 
SMGs (Greve et al. 2005; Carilli \& Wang 2006; Coppin et al. 2008a) 
and the z$\sim$1.5 massive star forming disk galaxies (Daddi et al. 2008). 
The molecular gas can provide the requisite fuel for massive star 
formation which is suggested by the FIR emission redshifted to  
submillimeter and millimeter wavelengths (Benford et al. 1999; 
Omont et al. 1996b, 2003; Priddey et al. 2003; 
Robson et al. 2004; Beelen et al. 2006; Wang et al. 2008a). 

The CO line emission detected in FIR luminous quasars provides estimates 
of the dynamical properties of the spheroidal bulges (i.e. bulge mass 
and velocity dispersion). Shields et al. (2006) investigated the relation between 
black hole mass ($\rm M_{BH}$) and bulge velocity 
dispersion ($\rm \sigma$) in high-z quasar systems using 
the observed CO line widths. They found that the massive 
quasars ($\rm M_{BH}>10^{9}$ to $\rm 10^{10}\,M_{\odot}$) at z$>$3 
appear to have much narrower $\rm \sigma$ values compared to 
what is expected from the local $\rm M_{BH}$--$\rm\sigma$ relationship 
(e.g., Tremaine et al. 2002). Moreover, Coppin et al. (2008a) found 
that the average black hole--bulge mass ratios for CO and FIR luminous 
quasar samples at z$\sim$2 are likely to be an order of magnitude 
higher than the local value ($\rm M_{BH}/M_{bulge}\sim0.0014$, Marconi \& Hunt 2003). 
Less evolved stellar bulges were also indicated by high resolution imaging of 
the CO emission in two z$>$4 quasars (Riechers et al. 2008a, 2008b). 
These results argue for the scenario that the formation of the SMBHs 
occurs prior to that of the stellar bulge, which is also suggested 
by the near-IR imaging of high-z quasar host galaxies (Peng et al. 2006a, 2006b).

CO line emission was previously searched in four z$\sim$6 quasars, 
and detected in two of these (Walter et al. 2003; Bertoldi et al. 2003b; 
Carilli et al. 2007; Maiolino et al. 2007). The two CO-detected quasars, 
J1148+5251 and J0927+2001, are the brightest millimeter sources 
among a sample of thirty-three z$\sim$6 quasars that have published 
millimeter dust continuum observations (Petric et al. 2003; Bertoldi et al. 2003a; 
Wang et al. 2007, 2008a). The detected CO transitions indicate highly 
excited molecular gas in the quasar host galaxies (line flux density spectral 
energy distributions peaking at \textit{J}=6 or higher, Carilli et al. 2007; 
Riechers et al. 2009) and the implied molecular gas masses are 
all $\rm \sim2\times10^{10}\,M_{\odot}$. High resolution VLA 
imaging has resolved the CO emission in J1148+5251 to kpc scale, suggesting 
a dynamical mass of $\rm M_{dyn}sin^2{\it i}\sim4.5\times10^{10}\,M_{\odot}$ 
within a radius of 2.5 kpc, where {\it i} is the inclination 
angle (Walter et al. 2004; Riechers et al. 2009) of the molecular disk.  
This provides the first direct constraint on the mass 
of the quasar host galaxy at the highest redshift, which 
suggests a high black hole--bulge mass 
ratio similar to that found with the z$\sim$2 and z$>$4 quasars.

In this work, we extend the CO observations to all z$\sim$6 quasars 
with known 250 GHz continuum flux densities 
of $\rm S_{250GHz}\geq1.8\,mJy$ (Wang et al. 2007, 2008b). We 
aim to study their general molecular gas properties 
and investigate possible constraints on the SMBH-host 
evolution with these earliest quasars.
We describe the sample selection and observations 
in Section 2, and present the results in Section 3. The CO emission and 
gas properties of these z$\sim$6 quasars are analyzed in Section 4. Based on 
our results, we present a brief discussion on the constraints of the black 
hole-bulge evolution in Section 5, and summarize the main conclusions in 
Section 6. A $\rm \Lambda$-CDM 
cosmology with $\rm H_{0}=71km\ s^{-1}\ Mpc^{-1}$, $\rm
\Omega_{M}=0.27$ and $\rm \Omega_{\Lambda}=0.73$ is adopted throughout this
paper (Spergel et al. 2007).

\section{Sample and observations}

\subsection{Sample}

There are thirty-three quasars at z$\sim$6 that have 
published observations of dust continuum at 250 GHz that 
were made with the Max-Planck Millimeter Bolometer 
Array (MAMBO, Kreysa et al. 1998) on the IRAM 30m-telescope, 
and ten of these have been detected (Bertoldi et al. 2003a; 
Petric et al. 2003; Willott et al. 2007; Wang et al. 2007, 2008b). 
The two brightest MAMBO detections (i.e. $\rm S_{250GHz}\sim5\,mJy$, 
$\rm L_{FIR}\sim10^{13}\,L_{\odot}$), J1148+5251 and J0927+2001, 
have all been detected in strong molecular CO line emission 
(J1148+5251: CO (3-2), Walter et al. 2003; 
CO (7-6) and (6-5), Bertoldi et al. 2003b; J0927+2001: 
CO (6-5) and (5-4), Carilli et al. 2007). 
In this work, we present new observations of CO (6-5)
and/or (5-4) line emission in another six z$\sim$6 
quasars (J0840+5624, J1044$-$0125, J1048+4637, J1335+3533, 
J1425+3254, J2054$-$0005), using the Plateau de Bure Interferometer (PdBI). 
They all have 250 GHz flux density 
of $\rm S_{250GHz}\geq1.8\,mJy$ and corresponding FIR 
luminosities of $\rm L_{FIR}>5\times10^{12}\,L_{\odot}$ (using
the estimation given in Wang et al. 2008a and 2008b). 
We also report our observations of the CO (2-1) line in J0840+5624 and
J0927+2001 with the Green Bank Telescope (GBT), CO (3-2) in J1048+4637 with
the Very Large Array (VLA), and 350 $\mu$m dust continuum
in J0840+5624, J0927+2001, J1044-0125, J1335+3533, 
and J1425+3254 with the SHARC-II bolometer
camera at the Caltech Submillimeter Observatory (CSO).

We list the basic information of all the eight 
FIR luminous z$\sim$6 quasars in Table 1, including
their redshifts from the discovery papers (Fan et al. 2000, 2003, 2006a,
Jiang et al. 2008, Cool et al.2006), optical magnitudes, continuum flux
densities at 250 GHz and 1.4 GHz (Bertoldi et al. 2003a; Carilli et al. 2004;
Wang et al. 2007; Wang et al. 2008b), as well as 
available measurements of the black hole masses and AGN bolometric 
luminosities (Jiang et al. 2006). These objects are mainly 
optically selected with rest-frame 1450$\rm \AA$ ($\rm m_{1450}$) 
absolute magnitude of $\rm -27.8 <M_{1450} < -26.0$, 
indicating AGN bolometric luminosity 
of $\rm L_{bol}>10^{13}\,L_{\odot}$ (Jiang et al. 2006), which are 
among the brightest objects in the quasar population. 
The black hole masses of J1148+5251 
and J1044$-$0125 have been determined from the 
measurements of the AGN UV broad line 
emission (Willotts et al. 2003; Jiang et al. 2006) 
which are a few $\rm 10^{9}\,M_{\odot}$, indicating 
Eddington ratios of 0.5 to 1.0 (Jiang et al. 2006). 

\subsection{Molecular CO observations}

The observations of the CO (6-5) and (5-4) lines were carried out with 
the new generation 3 mm receiver 
on the PdBI, which provides a bandwidth of 1 GHz 
in dual polarization. The sources were observed between 2007
and 2009, in the compact D configuration (FWHM resolution $\rm \sim5''$). 
We generally used two frequency setups for the first 
two tracks on each target to cover a continuous bandwidth 
of 1.8 GHz, to account for the possible uncertainties 
between the optical and systemic (CO) redshifts. In the case of marginal detections, 
we added one more track with both polarization bands centered at 
the likely line frequency to confirm the line at $\rm >3\sigma$ 
significance. 
The original frequency resolution is 2.5 MHz which was subsequently binned 
to 23 MHz ($\rm \sim 70\,km\,s^{-1}$). 
Phase calibration was performed every 20 min with observations 
 of 0954+658, 1055+018, 1150+497, 1308+326, 
and 2134+004. We observed MWC 349 as the 
flux calibrator. The typical rms after eight hours of observing time 
is about $\rm 0.5\,mJy\,beam^{-1}$ per $\rm 70\,km\,s^{-1}$ wide bin. The data 
were reduced with the IRAM GILDAS software package (Guilloteau \& Lucas 2000). 

We also conducted observations of CO (2-1) line emission 
in two of the CO-detected z$\sim$6 quasars, J0840+5624 and J0927+2001, 
during the winter months of 2007-2008, 
using the Ka-band receiver on the 100-m GBT. 
The observations employed the subreflector 
nodding mode with half-cycle times between 9.0 to 22.5 seconds. 
The tunings were based on the CO (6-5)/(5-4) redshifts 
 (Bertoldi et al. 2003b; Carilli et al. 2007). 
We set up the spectrometer in low resolution mode with 
two single-polarization 800 MHz windows centered at offsets of $\rm -100\,MHz$ 
and $\rm +100\,MHz$ from the observed line frequency. 
Thus, the two polarizations cover the redshifted CO (2-1) 
line with a bandwidth of 600 MHz simultaneously. 

Flux calibration was performed using 3C147, and pointing
was checked regularly on known radio sources. Data were 
analyzed using the data reduction routines in the GBTIDL 
software package. The uncertainties of the flux calibration 
are typically 10\% to 20\%. We have spent 17.7 hours of  
observing time (including overhead due to subreflector 
movement) on J0840+5624, leading to an rms noise of about 0.13 mJy 
per 50 km s$^{-1}$ wide channel, measured from the line-free channels. 
However, there is a small-scale baseline
structure close to the observed line frequency, 
which cannot be removed in the calibration, and thus, precluded 
detection of weak line emission with peak 
flux density of a few hundred $\rm \mu Jy$ (see Wagg et al. 2008). 
We also spent 10.7 hours on J0927+2001, and the CO (2-1) line is not detected 
with a 1$\sigma$ rms sensitivity of 0.15 mJy per 50 km s$^{-1}$ channel. 

Finally, we report a tentative VLA Q-band detection of CO (3-2) line emission from 
J1048+4637. These observations were conducted in
2004 in BC, C, and CD configurations. 
The source was observed in continuum mode, i.e. 
two polarizations, two Intermediate Frequency
bands "IFs", and 50 MHz bandwidth ($\sim$310 km s$^{-1}$ in velocity) 
per IF for each frequency setup. 
We searched the CO (3-2) line emission in the frequency 
range from 47.715 GHz to
48.165 GHz (corresponding to z=6.179 to 6.247 in redshift) 
by observing the source repeatedly with different frequency setups. 

We reduced the data and made images using the standard VLA wide field data
reduction software AIPS and the source is detected in 
the 50 MHz channel centered at 47.865 GHz (z=6.224, 
$\rm \Delta z=0.008$) at $\rm \gtrsim$4$\sigma$ 
with a 1$\sigma$ rms of 70 $\rm \mu Jy\,beam^{-1}$. 
The spatial resolution on the final map is 0.5$''$ (FWHM).
The 1$\sigma$ rms values for the other channels 
are much higher, i.e. from 120 to 200 $\rm \mu$Jy beam$\rm ^{-1}$ 
(due to less observing time) and no signal is 
detected at the quasar postion. We excluded the data observed 
at 47.865 GHz and 47.815 GHz which are close to the CO (3-2) 
line frequency for z=6.2284 (i.e. the redshift determined 
with the PdBI CO (6-5) detection), and merged the data of all 
the other channels to constrain the source continuum at 48 GHz. 
The 1$\rm \sigma$ rms noise on the combined image is 
65 $\rm \mu$Jy beam$\rm ^{-1}$ and no continuum 
emission is detected.

\subsection{SHARC-II 350 $\mu$m dust continuum observations}

We obtained new 350 $\mu$m observations of the dust continuum emission  
from five z$\sim$6 quasars, using the SHARC-II bolometer 
camera on the CSO 10.4 m telescope. The SHARC-II camera 
is a 12$\rm \times$32-element array with a field of 
view of $\rm 2.6'\times1.0'$ and a beam size of $\rm 8.5''$.
The observations were conducted in February 2008 during excellent 
weather conditions on Mauna Kea (opacity at 225 GHz $<$ 0.06). 
We adopted a scan pattern similar to our previous SHARC-II 
observations of z$\sim$6 quasars, i.e. a Lissajous patten 
with amplitudes of $\rm \pm45''$ and $\rm \pm12''$ in azimuth 
and elevation respectively. This provides a uniform coverage 
of $\sim65''\times34''$. Pointing, focus, and flux 
calibrations were done on Mars, Saturn, and a number 
of secondary calibrators (OH231.8+4.2, CRL618, IRC10216, Arp220, 
and 3C345). The final calibration uncertainties are within 20\%.

The on-source integration time for each of the five objects is listed 
in Table 2. Data reduction was performed with the CRUSH data reduction package
 version 1.63 (Kov$\rm\acute{a}$cs et al. 2006a) and the rms noise values in the final maps 
are $<$ 5 mJy beam$\rm ^{-1}$ for all targets. With these 
new observations, we confirm a previous 350 $\mu$m detection 
of J0927+2001 and marginally detect the dust continuum 
toward J0840+5624 ($\rm 2.9\sigma$). 
The measurements and upper limits for the other three sources are 
listed in the next section. 

\section{Results}  

We have detected CO (6-5) or (5-4) line emission at $\rm \geq5\sigma$ 
in all six quasars observed with the PdBI. 
The CO emission is unresolved for 
all the six objects with the 5$\rm ''$ synthesized 
beam ($\rm \sim29.2\,kpc$ at z$\sim$6), and only one 
source, J1048+4637, has a clear detection of the 3 mm continuum in  
the line-free channels. We fit the line spectra with a single Gaussian 
profile to determine the line widths (FWHM), host-galaxy redshifts, and 
corresponding uncertainties. The line spectra with the Gaussian-fit 
profiles are presented in the left panel of Figure 1. 
The line fluxes are obtained by integrating the data 
over the line emitting channels. The averaged-intensity maps 
are presented in the right panel of Figure 1. 
The CO emission peaks on the intensity maps show offsets 
of $\rm 0.3''$ to $\rm 1.8''$ from the optical quasar 
positions (see details for each sources below), which 
are all within the relative astrometric uncertainties of the 
observations. We summarize the line and continuum parameters 
for the PdBI detections in Table 2. The CO (3-2) detection 
in J1048+4637, CO (2-1) upper-limits, 
and the 350 $\mu$m dust continuum measurements are also listed in this table. 
We show the 350 $\mu$m dust continuum maps 
of J0840+5624 and J0927+2001 in Figure 2. 
The VLA Q-band 47.865 GHz image of the CO (3-2) line 
detection in J1048+4637 is shown in the 
left panel of Figure 3. An average image 
of the line-free channels is also 
presented (right panel), which constrains the 
continuum emission from this object at 48 GHz.

\subsection{Note for individual objects}

\hspace{-2.em}{\bf J0840+5624} 
Both the CO (6-5) and (5-4) transitions were 
detected, peaking at $\rm 08^{h}40^{m}35.08^{s}$, 
$\rm +56\degrees24'20.5''$, $\rm 0.6''$ from
the optical quasar position. 
The integrated line flux densities are 0.60$\pm$0.07 Jy km s$\rm^{-1}$ and 
0.72$\pm$0.15 Jy km s$\rm^{-1}$, respectively. 
A single Gaussian line profile fitted  
to the combined spectrum of the two CO transitions yields a host galaxy 
redshift of 5.8441$\pm$0.0013 and a FWHM line width
of 860$\pm$190 km s$\rm ^{-1}$. This is the largest 
CO line width that we have detected among the z$\sim$6 
quasars, and the line profile is double-peaked. 
A fit with two Gaussian components to the combined 
spectrum gives peak velocity offsets of $\rm -300\,km\,s^{-1}$ 
and $\rm 230\,km\,s^{-1}$, and FWHM line widths of $\sim$300 $\rm km\,s^{-1}$ 
and 410 $\rm km\,s^{-1}$ for the two components, respectively. 
Continuum is not detected in the line-free 
channels and the 3$\rm \sigma$ upper limits are 0.15 
mJy at 85 GHz and 0.30 mJy at 101 GHz. 

The GBT observation of the CO (2-1) 
transition gives an rms of about 0.13 mJy per 50 km s$\rm ^{-1}$ 
channel. 
Assuming a velocity range similar to the CO (5-4) line, 
We estimated a 3$\sigma$ line flux upper limit of about 0.1 Jy 
km s$\rm ^{-1}$ for this object\footnote{
The 3$\sigma$ upper limits on the CO (2-1) line emission 
are estimated as $\rm 3\,\sigma_{channel}(\Delta v_{channel}\,
\Delta v_{line})^{1/2}$ (in Jy km s$\rm ^{-1}$), where 
$\rm \Delta v_{line}$ is the expected CO line width 
in km s$\rm ^{-1}$, $\rm \Delta v_{channel}$ is 
the channel width in km s$\rm ^{-1}$, 
and $\rm \sigma_{channel}$ is the corresponding rms noise 
value in Jy (Seaquist et al. 1995; Wagg et al. 2007).
We adopt the full line width at zero intensity from the PdBI spectrum 
as $\rm \Delta v_{line}$ in the calculation.}. 
However, we emphasize that there are 
large uncertainties in this measurement due to a small-scale 
baseline structure close to the expected line frequency (Wagg et al. 2008). 

The 350 $\mu$m dust continuum emission is marginally 
detected in this object at $\rm \sim2.9\sigma$. 
The continuum source is un-resolved and peaks at 
RA=$\rm 08^{h}40^{m}35.32^{s}$ Dec=56d24$'$17.5$''$, 
which is 3.1$''$ away from the optical quasar (the left 
panel of figure 2). The peak value is 9.3$\pm$3.2 mJy 
beam$\rm ^{-1}$ on the final map and we adopt this as 
the tentative 350 $\mu$m flux density. Further 350 $\mu$m 
observations with better sensitivity are necessary
to confirm this tentative detection. On the SHARC-II 
map, there is also a 4$\sigma$ peak (14.8$\pm$3.7 mJy) 
at the position of RA=$\rm 08^{h}40^{m}34.35^{s}$ 
Dec=56d24$'$45.0$''$ (i.e. northwest to the quasar, see the left panel of Figure 2). 
We checked our previous VLA 1.4 GHz continuum 
observation (observed in A array with a resolution 
of FWHM$\rm\sim1.4''$, Wang et al. 2007) and there is a radio   
source with $\rm S_{1.4GHz}=44\pm9\,\mu Jy$ close 
to this position, i.e. at RA=$\rm 08^{h}40^{m}34.25^{s}$ Dec=56d24$'$44.7$''$. 
The offset is only $\rm 0.9''$, well within the position uncertainty of 
the SHARC-II observation. There is no further 
identification information for this field source yet. 

\hspace{-2.0em}{\bf J1044$-$0125} 
 We detected the CO (6-5) line in this 
source, peaking at the position 
of $\rm 10^{h}44^{m}33.02$, $\rm -01\degrees25'02.3''$, $\rm 0.3''$ 
away from the optical quasar. The FWHM of the CO 
line fitted with a single Gaussian profile 
is 160$\pm$60 km s$\rm ^{-1}$,  which represents the narrowest CO line 
width among all the CO-detected z$\sim$6 quasars. 
The integrated line flux is 0.21$\pm$0.04 Jy km $\rm s^{-1}$. The 3$\rm \sigma$ upper 
limit on the continuum emission is 0.15 mJy at 102 GHz. 

The 350 $\mu$m dust continuum emission has not been 
detected in our SHARC-II observations, and the 1$\rm\sigma$ rms 
is 4.5 mJy beam$\rm^{-1}$. This yields a $\rm 3\sigma$ flux density upper 
limit of 13.5 mJy. 

\hspace{-2.0em}{\bf J1048+4637}  
We detected the CO (6-5) line emission in this source, 
peaking at $\rm 10^{h}48^{m}45.08$, $\rm 46\degrees37'18.7''$, $\rm 0.5''$ away 
from the optical quasar position. 
The CO redshift is 6.2284$\pm$0.0017, 
which is close to the value of 6.23 
estimated from the Ly$\rm \alpha +$NV emission in the 
quasar rest-frame UV spectrum (Fan et al. 2003), 
but higher than the value of 6.19$\sim$6.20 measured 
from the Mg II $\rm \lambda 2798\AA$ emission 
(Iwamuro et al. 2004; Maiolino et al. 2004). 
The 3~mm dust continuum emission is 
detected along with the CO line emission.
A simultaneous fit of the CO lines and the 
continuum gives a continuum flux density of $\rm 0.13\pm0.03\,mJy$ at 96 GHz. 
The integrated line flux is 0.27$\pm$0.05 Jy km s$\rm ^{-1}$, and 
the FWHM line width from single Gaussian fit is 370$\pm$130 km s$\rm ^{-1}$.
We have also detected CO (6-5) line emission
from a source $\rm \sim28''$ away, to the northeast of the
quasar (J1048NE). The CO spectrum yields a redshift of 6.2259$\pm$0.0019 for this
NE source, indicating the presence of a companion galaxy $\rm \sim160$ kpc
away from the optical quasar.  
The observed peak flux density  
is 0.53$\pm$0.14 mJy, which gives an integrated flux 
of 0.23$\pm$0.05 Jy km s$\rm ^{-1}$. Considering an attenuation 
factor of $\sim0.4$ for the primary beam ($\rm FWHM\sim52''$) 
at the source position, we estimate the intrinsic line flux 
to be 0.58$\pm$0.13 Jy km s$\rm ^{-1}$ (see Figure 1). 

The CO (3-2) line emission from this object was searched  
using the Q-band receiver on the VLA, and we detected the 
source at $\rm 0.30\pm0.07\,mJy$ in a 50 MHz wide channel 
centered at 47.865 GHz. The data averaged over the other 
channels present a 3$\rm \sigma$ upper limit of $\rm 0.195\,mJy$ 
for the continuum emission at 48 GHz, indicating that the 
$\rm 0.3\,mJy$ detection at 47.865 GHz are not due 
to nonthermal quasar contniuum emission. An 
optically thin graybody model with a dust temperature 
of $\rm T_{dust}=47 K$ and an emissivity index of $\rm \beta=1.6$ 
(Beelen et al. 2006; Wang et al. 2008b) fitted to the continuum 
measurements at 250 GHz and 96GHz suggests a continuum 
flux density of only 0.017 mJy at the observed frequency of 47.865 GHz. 
Thus the $\rm 0.30\,mJy$ detection in this channel 
are likely to be from the CO (3-2) line emission. 
The source on the VLA image is marginally resolved 
by the $\rm 0''.5\times0''.5$ synthesized beam, and a fit 
to a 2D-Gaussian component yields a source size 
of $\rm 0''.9\times0''.4$. 
The integrated line intensity detected in 
the channel is $\rm 0.09\pm0.02\,Jy\,km\,s^{-1}$. 
According to the line profile and source redshift 
of z=6.2284$\pm$0.0017 determined from
the CO (6-5) transition, the VLA observation is likely to 
be centered at the blue part of the CO (3-2) line emission 
and covers a bandwidth of 310 km s$^{-1}$ in velocity.
Thus we adopt the measurement as a lower limit of 
the CO (3-2) line flux for this object. 


\hspace{-2.0em}{\bf J1335+3533} 
The CO (6-5) line was detected in this source, peaking at the
position of $\rm 13^{h}35^{m}50.75^{s}$, $\rm 35\degrees33'15.8''$, 
$\rm 0.7''$ away from the optical quasar position.
The CO FWHM fitted with a single Gaussian profile 
is 310$\pm$50 km s$\rm ^{-1}$. 
The integrated line intensity
is 0.53$\pm$0.07 Jy km s$\rm ^{-1}$, and 
the 3$\rm \sigma$ upper limit on the continuum emission is 0.15 mJy at 100 GHz.

No 350 $\mu$m dust continuum is detected in our SHARC-II observations
of this object. The 1$\sigma$ rms on the final 
map is 4.4 mJy beam$\rm^{-1}$, and the corresponding 3$\rm\sigma$ 
upper limit is 13.2 mJy for the 350 $\mu$m flux density. 

\hspace{-2.0em}{\bf J1425+3254} 
The CO (6-5) line was detected, peaking at the
position of $\rm 14^{h}25^{m}16.26^{s}$, $\rm 32\degrees54'09.3''$, 
$\rm 0.6''$ away from the optical quasar.
The CO FWHM fitted with a single Gaussian profile 
is 690$\pm$180 km s$\rm ^{-1}$, and the line flux 
of 0.59$\pm$0.11 Jy km s$\rm ^{-1}$.
The 3$\rm \sigma$ upper limit for the continuum at 101 GHz is 0.12 mJy.

This source was not detected in our SHARC-II observations 
of the 350 $\mu$m dust continuum with a 1$\sigma$ rms of 2.6 mJy beam$\rm^{-1}$. 
This yields a 3$\sigma$ upper limit of 7.8 
mJy for the 350 $\mu$m flux density. 

\hspace{-2.0em}{\bf J2054$-$0005} 
The CO (6-5) line was detected in this source, peaking at the
position of $\rm 20^{h}54^{m}06.45^{s}$, $\rm -00\degrees05'13.1''$, $\rm 1.8''$ 
away from the optical quasar.
The CO line FWHM obtained from a fit to a single-Gaussian line profile is 360$\pm$110 km s$\rm ^{-1}$. 
The derived integrated intensity 
is 0.34$\pm$0.07 Jy km s$\rm ^{-1}$, and the 3$\rm \sigma$ upper limit on
the continuum emission at 98 GHz is 0.15 mJy. 

\hspace{-2.0em}{\bf J0927+2001} 
We have observed the CO (2-1) line in this object for 10.7
hours with the GBT, and the final rms is 0.15 mJy
per 50 km s$\rm ^{-1}$ channel. 
We adopt a line width of $\sim$900 km s$\rm ^{-1}$ (full 
width at zero intensity) based on the previous PdBI 
detections of the CO (6-5) and (5-4) 
transitions (Carilli et al. 2007), and derive a 3$\sigma$
upper-limit to the integrated line intensity of $\sim$0.1 Jy km s$\rm ^{-1}$.

The 350 $\mu$m dust continuum emission from this object was previously
detected by our SHARC-II observations in 2007 (17.7$\pm$5.7 mJy, Wang et al. 2008a). 
The new data we presented here confirms
the detection with better sensitivity, i.e. a 350 $\mu$m flux density
of 11.7$\pm$2.4 mJy. This 350 $\mu$m dust
continuum peak is quite consistent with the optical quasar
position (offset of $\rm <1''$). We cannot confirm
the $\rm \sim3\sigma$ peak 15$''$ away from the quasar
position reported in our previous work (Wang et al. 2008a).

\subsection{CO redshifts}

We compare the redshifts determined from
CO and UV emission lines in Table 3 for all CO detected z$\sim$6 quasars,
including the previous CO-detections in J1148+5251 and J0927+2001
(Bertoldi et al. 2003b, Carilli et al. 2007). The C {\scriptsize IV} $\rm \lambda 1549\AA$
redshifts of J0840+5624 and J1044$-$0125 are blue-shifted 
by about $\rm 3000$ and $\rm 1600\,km\,s^{-1}$ compared to
the values determined from the CO lines, respectively. This is consistent with
the fact that the C {\scriptsize IV} emission is typically blue-shifted from the quasar systematic redshift
by up to a few thousand $\rm km\,s^{-1}$ (Richards et al. 2002; Goodrich et al. 2001; Ryan-Weber et al. 2009).
There is no systematic offset between the CO redshifts and those determined from other UV
emission lines. In particular, the CO redshifts of J1044$-$0125 and J1148+5251
are in good agreement with the values derived from the C{\scriptsize III]} $\rm \lambda 1909\AA$
and Si{\scriptsize IV} $\rm\lambda 1400\AA$ broad emission lines with high-quality near-infrared
spectroscopy (Jiang et al. 2007; Ryan-Weber et al. 2009). The largest offset
between CO and Ly$\alpha$ redshifts is $\rm \sim$0.04, which is reasonable since the Ly$\alpha$
redshifts are always poorly constrained (with uncertainties of 0.02$\sim$0.05)\footnote{Note that
the Ly$\alpha$ redshift uncertainty of 0.004 quoted in Jiang et al. (2008) is only an
error from their spectral fitting.}.
Three objects have redshifts determined from the 
Mg {\small II} $\rm \lambda 2798\AA$ line. The CO redshifts of J1044$-$0125 
and J1148+5251 are consistent with the Mg {\small II} measurements 
within the uncertainties, while a large offset of $\rm \sim0.03$ 
is seen in J1048+4637 between the CO redshift and the Mg {\small II} ones 
quoted in Iwamuro et al. (2004) and Maiolino et al. (2004). 

\section{Analysis}

Our PdBI observations have a 100\% CO detection rate 
among the millimeter (FIR) luminous quasars in the early universe. 
Together with the previous CO detections in J1148+5251 and 
J0927+2001 (Bertoldi et al. 2003b; Walter et al. 2003, 2004; 
Carilli et al. 2007), all eight z$\sim$6 quasars with 
published 250 GHz flux densities of $\rm S_{250GHz}\ge1.8\,mJy$ 
(Bertoldi et al. 2003a; Wang et al. 2007; Wang et al. 2008b)
have now been detected in CO (6-5) and/or (5-4) line emission.
In this section, we present a study of the CO emission line properties 
in the earliest quasar host galaxies with this sample. We compare 
their molecular gas masses and CO line width distributions to that 
of the CO-observed quasars and submillimeter galaxies at lower 
redshift. A list of fifteen CO detected quasars at $\rm 1.4\leq z\leq4.7$
was summarized in SV05. 
Another quasar SDSS J0338+0021 at z=5.0267 was detected 
in CO (5-4) line emission by Maiolino et al. (2007). Additionally, 
Coppin et al. (2008a) presented six CO-detected quasars at 
$\rm 1.7\leq z\leq2.6$ (one 
of these is also in the SV05's sample). These give a
sample of twenty-one CO-detected quasars with redshifts from 1.4 to 5. 
We also consider the a sample of fourteen CO-detected SMGs at $\rm 1.1\leq z\leq 3.4$
from Greve et al. (2005) and Tacconi et al. (2006).

\subsection{Molecular gas masses}

The mass of the molecular gas in J1148+5251 is well determined from
the CO (3-2) data (Walter et al. 2003, 2004; Riechers et al. 2009). 
The high-order CO transitions ($\rm J\geq5$) detected in SMGs 
and high-redshift quasars are usually not thermalized (e.g., Wei$\ss$ et al. 2005a, 2007; 
Riechers et al. 2009). Thus, for the other seven CO-detected z$\sim$6 quasars, 
We adopt the line ratios of $\rm L'_{CO(6-5)}/L'_{CO(1-0)}\approx0.78$
and $\rm L'_{CO(5-4)}/L'_{CO(1-0)}\approx0.88$ from the multi CO transition large
velocity gradient (LVG) modeling of J1148+5251 (Riechers et al. 2009) 
and use these numbers to calculate the CO (1-0)
line luminosities ($\rm L'_{CO(1-0)}$). We then estimate the molecular gas
masses ($\rm M_{gas}=M[H_2+He]$) within the CO emitting region of the quasar host galaxies
using $\rm M_{gas}=\alpha L'_{CO(1-0)}$. An integrated CO intensity-to-gas
mass conversion factor of $\rm \alpha=0.8 M_{\odot}\,(K\,km\,s^{-1}\,pc^{2})^{-1}$, 
appropriate for ULIRGs is adopted here (Solomon et al. 1997;
Downes \& Solomon 1998; SV05). 
The derived molecular gas masses are in the range, 
$\rm 0.7\times10^{10}$ to $\rm 2.5\times10^{10}\,M_{\odot}$ (Table 4) 
with a median value of $\rm 1.8\times10^{10}\,M_{\odot}$. 

We plot the molecular gas mass distribution of the z$\sim$6 quasars,
CO-observed SMGs, and $\rm 1.4\leq z\leq5$ quasars in the left
panel of Figure 4. The range of gas masses of the z$\sim$6 quasars 
($\rm 0.7\times10^{10}$ to $\rm 2.5\times10^{10}\,M_{\odot}$) appears narrower 
but still comparable to the typical values found in the other two samples.
We performed the standard Kolmogorov-Smirnov test between the z$\sim$6 
quasars and the other two samples. The probabilities that 
the data are drawn from the same parent population are 19\% for the z$\sim$6 
and $\rm 1.4\leq z\leq5$ quasar samples, and 16\% for the z$\sim$6 quasar and SMG samples. 
This agrees with the results from previous studies 
that the high-z FIR and CO luminous quasars show molecular gas mass 
distribution similar to that of the SMGs (Carilli et al. 2006; Coppin et al. 2008a).

\subsection{CO line widths}

The CO line widths (FWHM) of the z$\sim$6 quasar sample are spread over a wide
range, from $\rm 160\,km\,s^{-1}$ to $\rm 860\,km\,s^{-1}$, with a median
value\footnote{The median value is the average
of the two middle values if an even number of data points 
are presented in the sample.} of $\rm 360\,km\,s^{-1}$.
The broadest CO line detection (J0840+5624)
appears to be double-peaked with peak-velocity offsets
of $\rm \sim270\,km\,s^{-1}$ (see Figure 1).
Similar line profiles are widely observed in samples 
of SMGs (Greve et al. 2005; Wei$\ss$ et al. 2005a; 
Tacconi et al. 2006), which may be due to either
uncoalesced molecular gas components in a galaxy 
merger or simply reflect a large inclination angle of an  
extended gas disk relative to the sky plane.

In the right panel of Figure 4, we compare the CO line-width
distribution of the z$\sim$6 quasar sample to that of the
CO-detected SMGs (Greve et al. 2005; Tacconi et al. 2006)
and $\rm 1.7\leq z\leq5$ quasars (SV05;
Coppin et al. 2008a). The Kolmogorov-Smirnov tests return probabilities 
of 13\% for z$\sim$6 quasars and SMGs, and 71\% for z$\sim$6 
and $\rm 1.7\leq z\leq5$ quasars, i.e. there is no
significant difference in the line-width distributions. This result
confirms that the observed line-width distribution of high-z CO and FIR quasars
is comparable to that of the SMGs as was found in Coppin et al. (2008a).
The systemic difference reported in previous 
works (Greve et al. 2005; Carilli \& Wang 2006) is likely to be
due to selection effects and small sample size. 

\subsection{Molecular CO excitation}

The detection of CO (6-5) line emission in all eight z$\sim$6 quasars
in our sample, 
implies highly excited molecular gas in their host galaxies.
Four of the CO-detected z$\sim$6 quasars, J0840+5624, J0927+2001,
J1048+4637, and J1148+5251, have observations of multiple CO line 
 transitions (Bertoldi et al. 2003b;
Carilli et al. 2007; this work). The CO "excitation ladder" of
J1148+5251 has been studied by Bertoldi et al. (2003b) 
and Riechers et al. (2009), and
the best LVG model fitted to the data suggests CO emission from a
single gas component with kinetic temperature of $\rm T_{kin}=50\,K$
and a molecular gas density of $\rm \rho _{gas}(H_2)=10^{4.2}\,cm^{-3}$.
Similar CO excitation conditions are also found in
the CO and FIR luminous quasar BR 1202-0725 
at z=4.69 (Carilli et al. 2002; Riechers et al. 2006) and 
some nearby starburst galaxies (e.g., the high excitation  
component in M82, Wei$\ss$ et al. 2005b).  
In Figure 5, we plot the CO
 excitation ladders of J0840+5624, J0927+2001, and J1048+4637, together with
the LVG models of J1148+5251, BR 1202-0725, and the high excitation component in M82. 
The line intensity ratios of $\rm I_{CO(5-4)}/I_{CO(6-5)}$ 
for J0840+5624 and lower limit of $\rm I_{CO(3-2)}/I_{CO(6-5)}$ for J1048+4637 
are consistent with the model values. 
J0927+2001 shows a lower value of $\rm I_{CO(5-4)}/I_{CO(6-5)}$ 
compared to the model, which, however, is still marginally consistent given
the large uncertainties in the measurements of both transitions. 
These results suggest that such a warm,
highly excited, single component gas model is a reasonable description of
the molecular gas in all of these quasar host galaxies. 
Further observations of additional CO transitions in the  
CO-detected z$\sim$6 quasars will address how the CO excitation properties
vary among these objects.

\subsection{FIR-to-CO luminosity relationship}

The FIR luminosities, $\rm L_{FIR}$, of the eight CO-detected
quasars at z$\sim$6 are derived using the submillimeter and millimeter
continuum measurements from the literature and the CO observations
in this work (Bertoldi et al. 2003a; Beelen et al. 2006;
Wang et al. 2007; 2008a, 2008b; Carilli et al. 2007), as 
listed in Table 4. A dust emissivity index of $\rm \beta=1.6$ (Beelen et al. 2006) is 
adopted in these calculations. J1148+5251 has a fitted dust
temperature, $\rm T_{dust}$, of 56 K (Beelen et al. 2006) and
J0927+2001 has $\rm T_{dust}=46\,K$, from fits to the submillimeter 
and millimeter measurements (this work; Wang et al. 2007, 2008a; 
Carilli et al. 2007; Riechers et al. 2009).
We adopt the average dust temprature of 47 K found in high-z quasar host
galaxies (Beelen et al. 2006) for the other six objects (see Wang et al. 2008b).

We plot $\rm L_{FIR}$ versus $\rm L'_{CO(1-0)}$ for the eight z$\sim$6 quasars
in Figure 6, and compare with local spiral, infrared luminous 
galaxies (LIRGs, Gao \& Solomon 2004),
ULIRGs (Solomon et al. 1997), the $\rm 1.4\leq z\leq5$ quasar and SMG samples
discussed above. For the high-z samples, $\rm L_{FIR}$ is re-calculated using 
(sub)mm measurements from the literature (Smail et al. 2002;
Kov$\rm \acute{a}$cs et al. 2006b; Chapman et al. 2005; Omont et al. 2003; Benford et al. 1999;
Beelen et al. 2006), using the $\rm \beta$ value mentioned above.
We assume $\rm T_{dust}=47\,K$ for the objects with
less than three (sub)mm measurements in the quasar sample (Beelen et al. 2006), while $\rm T_{dust}=35\,K$
is used for the SMGs, which is typical for the $\rm T_{dust}$ values
found in previous (sub)mm studies (Kov$\rm \acute{a}$cs et al. 2006b; 
Coppin et al. 2008b). The z$\sim$6 quasars  
fall along the trend defined by all the other samples  
systems within the scatter (see the open stars in Figure 6). 

A fit to the samples of local spirals, LIRGs, ULIRGs, 
and SMGs using the Ordinary Least Squares Bisector 
method (Isobe et al. 1990) yields a relationship 
of $\rm log\,L_{FIR}=1.67\times log\,L'_{CO(1-0)}-4.87$ 
for these typical star-forming systems, which is 
consistent with the result of  
$\rm log\,L_{FIR}=1.7\times log\,L'_{CO(1-0)}-5$ quoted in SV05 with similar samples.
We also notice that the z$\sim$6 quasars all lie above this relationship. 
There are a number of undetermined parameters for 
individual objects, e.g., unknown AGN contributions to the FIR emission, different 
dust temperatures and CO line ratios, which may account 
for the offsets and scatters. Further observations at infrared, submillimeter, and millimeter 
wavelengths will be important to directly probe 
the infrared SEDs and better constrain the dust 
temperatures and AGN contributions for these objects. 
For a rough estimation, we derive  
the AGN contributions to the FIR emission for our sample with the quasar 
rest-frame 1450$\AA$ continuum (Table 1) and a FIR-to-1450$\AA$ 
luminosity ratio of $\rm L_{FIR}/\nu L_{\nu,1450\AA}=0.14$ 
from the local radio quiet quasar template (Elvis et al. 1994). 
The estimated AGN-dominated FIR emission is 
$\rm 0.9$--$4.4\times10^{12}\,L{\odot}$ for the eight 
sources, which accout for about 30\% of 
the FIR luminosities we calculated above with the (sub)mm 
observations on the average. We subtract these from the 
original $\rm L_{FIR}$ values (Table 1) and the corrected data points 
(see the filled stars in Figure 6) show a better agreement with 
the FIR-to-CO luminosity relationship for star-forming system.

\section{Discussion}

\subsection{Star forming activity in the millimeter and CO-detected quasars at z$\sim$6}

CO emission has been detected toward the eight z$\sim$6 quasars 
with published 250 GHz dust continuum flux densities 
of $\rm S_{250GHz}\geq1.8\,mJy$, indicating $\rm \gtrsim10^{10}\,M_{\odot}$ 
of molecular gas in the host galaxies of the FIR luminous quasars at the 
earliest epoch. The derived FIR and CO (1-0) luminosities     
follow the $\rm L_{FIR}-L'_{CO}$ relationship defined by 
star forming systems at low and high redshifts, suggesting a star-formation 
origin of a dominant fraction of the FIR emission in most of these objects, i.e. 
the excess FIR dust emission is powered by star 
forming activity in the circumnuclear region, not the central 
AGN (Bertoldi et al. 2003a; Wang et al. 2008b; Walter et al. 2009).
This is consistent with the idea of co-eval star formation with 
SMBH accretion in the massive z$\sim$6 
quasar systems which are bright at both UV-optical 
and FIR wavelengths (eg. Bertoldi et al. 2003a, 2003b; 
Wang et al. 2007, 2008a, 2008b; Walter et al. 2003, 2004, 2009). 

The measurements of (sub)mm dust continuum and molecular CO 
line emission from the quasar host galaxies provide 
the first constraint on the star forming activities in these objects. 
Using the AGN contribution-removed FIR luminosities derived in the 
previous section (Table 1), we estimate the star formation 
rates (SFRs)\footnote{Calculated with
$\rm SFR\sim 1.7\times10^{-10}L_{IR}\,\,(M_{\odot}yr^{-1})$,
where $\rm L_{IR}$ is the infrared luminosity (8-1000$\,\mu$m) in units
of $\rm L_{\odot}$. We assume $\rm L_{IR}$ 
is $\rm\sim 1.5L_{FIR}$ for 40$\sim$60 K warm dust emission.} 
for the eight z$\sim$6 quasars to be $\rm \sim530$ 
to $\rm 2380\,M_{\odot}\,yr^{-1}$ (Kennicutt 1998). 
The ratios between SFR and $\rm M_{gas}$ (i.e. SFR per 
solar mass of molecular gas) provide a 
measure of the star formation efficiency, which 
are about $\rm 5\times10^{-8}\,yr^{-1}$ 
to $\rm 1\times10^{-7}\,yr^{-1}$. 
This is comparable to that 
of the SMGs, $1.4\leq z\leq5$ CO-detected quasars, and local 
ULIRGs (SV05), and systematically higher than that of 
the local normal spiral disks (Kennicutt 1998), 
and high-z star-forming disk galaxies (e.g., 
the z$\sim$1.5 BzK galaxies, Daddi et al. 2008, 2009), suggesting 
a high star formation efficiency in the host galaxies of these 
FIR and CO luminous z$\sim$6 quasars similar to that 
of the extreme starburst systems (Tacconi et al. 2006; Walter et al. 2009). 
The inverse gives the gas depletion time scales 
of $\rm \tau_{dep}=M_{gas}/SFR\sim1$--$\rm 2\times10^{7}\,yr$. 

\subsection{Black hole--bulge evolution}

Massive bursts of star formation are likely to be on-going in these FIR
and CO bright quasars at z$\sim$6. 
In this section, we investigate the possible constraints of 
the black hole-bulge masses and velocity dispersion relationships 
in these z$\sim$6 quasar systems using the available 
CO measurements. The available SMBH masses and AGN 
bolometric luminosities ($\rm L_{bol}$) 
from the literature are listed in Table 1. 
For other objects, we adopt the $\rm L_{bol}$ values estimated 
in Wang et al. (2008b) from $\rm M_{1450}$, corrected to 
the CO redshifts, and estimated the black hole masses 
assuming $\rm L_{bol}/L_{EDD}=1$ and 
$\rm M_{BH}=L_{bol} (ergs\,s^{-1})/1.26\times10^{38}\,M_{\odot}$. 
The estimated $\rm M_{BH}$ values are all of the order of $\rm 10^{9}\,M_{\odot}$ 

\subsubsection{$\rm M_{BH}$--$\rm M_{bulge}$ relationship}

The dynamical masses ($\rm M_{dyn}$) of the eight z$\sim$6 quasars 
can be expressed as $\rm M_{dyn}\approx2.3\times10^5{v_{cir}}^2R$ 
(e.g., Neri et al. 2003; Walter et al. 2004; 
SV05; Narayanan et al. 2008), where $R$ is
the disk radius in kpc and $\rm v_{cir}$ is
the maximum circular velocity of the gas disk in km s$^{-1}$. 
High-resolution CO observations of J1148+5251 give a
gas disk radius of $\sim$2.5 kpc and a dynamical mass of 
4.5$\rm \times10^{10}\,M_{\odot}$. 
In the absence of spatially resolved measurement 
for our sample quasars, we here assume a disk radius 
of 2.5 kpc also for the other seven objects, and estimate $\rm v_{cir}$ 
using 1/2 of the full width at 20\% maximum (Ho 2007a, 2007b), which  
corresponds to about 3/4 of the CO FWHM for a single-Gaussian 
profile i.e. $\rm v_{cir}=3FWHM/4sin{\it i}$, where $i$ is 
the inclination angle of the molecular gas disk. We leave sin{\it i} 
as an unknown factor in these calculations. The derived 
$\rm M_{dyn}\,sin^2{\it i}$ are from $\rm 8.4\times10^{9}$ 
to $\rm 2.4\times10^{11}\,M_{\odot}$ (Table 4). For 0840+5624, if we 
adopt the average peak-velocity offset of 270 $\rm km\,s^{-1}$ 
as the disk circular velocity, the $\rm M_{dyn}\,sin^2{\it i}$ 
will reduce from $\rm 2.4\times10^{11}\,M_{\odot}$ to $\rm 4.2\times10^{10}\,M_{\odot}$. 

We then estimate the masses of the stellar components ($\rm M_{bulge}$) in the
spheroidal bulges as $\rm M_{bulge}=M_{dyn}-M_{gas}$. 
We adopt a series of inclination angle values, i.e. 
i=1$\degrees$, 5$\degrees$, 10$\degrees$, 20$\degrees$, 
30$\degrees$, 40$\degrees$, 50$\degrees$, 60$\degrees$, and 90$\degrees$, 
and derive $\rm M_{dyn}$ and $\rm M_{bulge}$ accordingly. 
The corresponding black hole-bulge mass ratios ($\rm M_{BH}/M_{bulge}$) 
versus different inclination angles are ploted in Figure 7, compared to the local
mass relationship of $\rm M_{BH}\sim0.0014M_{bulge}$.
The plot shows that if the sources were falling on the local black 
hole-bulge mass ratio of 0.0014 this would require inclination 
angles from $\rm <5\degrees$ to $\rm \sim25\degrees$, 
or $\rm <5\degrees$ to $\rm 15\degrees$ 
when the two broadest line objects are excluded. 

However, the observed distribution of CO line widths of the z$\sim$6 
quasar sample is similar to that of the SMGs, which argues
against systematically smaller inclination angle values. 
Moreover, these z$\sim$6 quasars 
are all among the brightest objects in the quasar population 
with AGN bolometric luminosities of $\rm L_{bol}\geq10^{13}\,L_{\odot}$. 
Recent studies showed that the unobscured fraction of quasars is 
increasing with quasar luminosity, and for objects with $\rm L_{bol}\geq10^{13}\,L_{\odot}$, 
this fraction is probably larger than 50\% (Simpson 2005; Treister et al. 2008). 
This is consistent with the receding torus model and 
implies a very large opening angle of the unobscured 
cone in these most luminous quasars, i.e. the central AGNs can 
be directly seen over an inclination angel range of {\it i} 
from $\rm 0\degrees$ to probably $\rm \geq60\degrees$ (Elitzur 2008).  
Thus, though we cannot rule out small inclination angle values
for individual objects, it is highly unlikely that more than half of these CO 
detected z$\sim$6 quasars are viewed in the extreme inclination 
angle range of $\rm i<15\degrees$. The CO estimated 
bulge dynamical masses probably reflect intrinsically 
smaller $\rm M_{BH}/M_{bulge}$ for these z$\sim$6 quasars 
compared to that of the local mature galaxies.

We emphasize that accurate calculations of the bulge 
dynamical and stellar masses still require 
high-resolution measurements of the disk size and geometry 
for each object. To roughly constrain the average 
black hole-bulge mass ratio of these FIR and CO 
luminous z$\sim$6 quasars here, we simply adopt an average 
inclination angle of $\rm 40\degrees$ based on the assumption of 
uniformly distributed {\it i} between $\rm 0\degrees$ and $\rm 60\degrees$. 
This results in a median $\rm M_{BH}/M_{bulge}$ ratio of 0.022 for 
our sample, which is about fifteen times higher than the present day value of 0.0014. 
This is in good agreement with the picture suggested 
by other high-z $\rm M_{BH}-M_{bulge}$ studies (e.g., Walter et al. 2004; 
Coppin et al. 2008a; Riechers et al. 2008a, 2008b; Peng et al. 2006a, 2006b), 
i.e. that the SMBH accumulates most its mass before the formation 
of the stellar bulge. 

For further consideration, if these z$\sim$6 quasars will finally evolve
into systems with black hole-bulge relationships identical to that in the
local universe, the final bulge stellar mass should be
around $\rm 10^{12}\,M_{\odot}$ with black hole masses on the order
of $\rm 10^{9}\,M_{\odot}$. The $\rm M_{BH}/M_{bulge}$ ratio 
we derived above suggests that only $\lesssim$10\% of the stellar component 
has been formed in these FIR and CO bright z$\sim$6 quasars on the 
average. On the other hand, the detected molecular gas mass
of $\rm \sim10^{10}\,M_{\odot}$ in these objects can only account
for $\rm <3\%$ of the mature bulge mass if the gas will all be converted to
stars. Thus, large amounts of gas supplied from external sources is 
required to form the $\rm 10^{12}\,M_{\odot}$ stellar bulge by z=0. 

\subsubsection{$\rm M_{BH}$--$\rm \sigma$ relationship}

We investigate the black hole mass-bulge velocity dispersion
relation with the eight CO-detected z$\sim$6 quasars in Figure 8.
The supermassive black hole mass ($\rm M_{BH}$) and bulge velocity
dispersion ($\rm \sigma$) in local galaxies follow a relationship
of $\rm log\,(M_{BH}/M_{\odot})=8.13+4.02log(\sigma/200\,km\,s^{-1})$ 
(Tremaine et al. 2002). We first roughly estimate $\rm \sigma$ for 
the eight z$\sim$6 quasars with the CO
line widths, using the empirical relation $\rm \sigma\approx FWHM/2.35$ 
(Shields et al. 2006; Nelson et al. 2000). 
The results are plotted in Figure 8, together with the local active 
and inactive galaxies from Tremaine et al. (2002) and CO-detected 
$\rm 1.4\leq z\leq5$ quasars that have available SMBH 
mass measurements (Shields et al. 2006; Coppin et al. 2008a). 
Most of the high-z CO-detected quasars, especially the objects 
with $\rm M_{BH}\geq10^{9}\,M_{\odot}$, are above the local $\rm M_{BH}-\sigma$ 
relationship with offsets of more than one order of 
magnitude in black hole mass, as was found in Shields et al. (2006). The median 
value of the ratios between $\rm M_{BH}$ and the 
expected black hole masses from the local $\rm M_{BH}$--$\rm\sigma$
relation ($\rm M_{BH,\sigma}$) for our z$\sim$6 
sample is $\rm \left <M_{BH}/M_{BH,\sigma}\right >_{median}\sim40$. 

This is consistent with the high $\rm M_{BH}/M_{bulge}$ ratios 
and the idea of prior SMBH formations we discussed 
in the previous section (\S5.2.1). However, the systemic 
offset between the high-z objects and the local $\rm M_{BH}-\sigma$
relation (i.e. $\rm \left <M_{BH}/M_{BH,\sigma}\right >_{median}$) 
can be reduced if we include assumptions 
of inclination angles in our estimations of $\sigma$,   
following the empirical correlations between $\sigma$ 
and inclination angle-corrected CO line width 
from the literature (eg. Ho 2007a, 2007b, Wu 2007).
In particular, if we consider the possible gas disk 
inclination angle range for these z$\sim$6 quasars 
and adopt an average value of $\rm i=40\degrees$ (
as was discussed in \S5.2.1), the derived 
$\rm \left <M_{BH}/M_{BH,\sigma}\right >_{median}$
will reduce to $\sim$26 and 4 following the methods 
described in Wu (2007) and Ho (2007a), respectively. 

We also notice that our sample exhibits significant 
scatter on the $\rm M_{BH}$--$\rm\sigma$ plot with 
derived $\rm M_{BH}/M_{BH,\sigma}$ values over three 
orders of magnitude. This is mainly due to the 
intrinsic scatter and offset between the real bulge 
velocity dispersions and the CO estimated values 
based on the empirical relationships of low-z samples. 
Indeed, how the observed CO line widths
trace the bulge velocity dispersions in
the high-z CO-detected quasars is still an open question. 
These objects are the most massive SMBH-host systems in the universe with SMBH
masses of a few $10^{9}$ to $\rm 10^{10}\,M_{\odot}$ and
experiencing massive star formation in the central kpc
scale regions (Walter et al. 2009). We will 
expect the sensitive interferometer 
arrays (such as ALMA and the EVLA) to directly probe the gas 
properties e.g.,
the gas distribution, geometry, 
dynamics (virialized or not), 
line profiles, and contributions from 
turbulence, etc. in their spheroidal hosts and finally address
the $\rm M_{BH}-\sigma$ correlation in these
high-z FIR and CO luminous quasars. 

\section{Conclusions}

We present new observations of molecular CO line emission and 350 $\mu$m dust 
continuum emission in quasar host galaxies at z$\sim$6. 
Our most important finding is that high-order 
CO transitions are detected in all six of the z$\sim$6 quasars
observed with the 3 mm receiver on the PdBI. The new CO detections all 
have observed 250 GHz dust continuum flux densities 
of $\rm S_{250GHz}\geq1.8\,mJy$. These results, together with previous CO-detections 
in another two objects, reveal an extremely high CO detection 
rate in the FIR luminous quasars at z$\sim$6. With the final sample 
of eight CO-detected z$\sim$6 quasars, we study the molecular gas 
properties in the earliest quasar host galaxies, and the main results 
are summarized as follows:

The CO emission indicates molecular gas masses of 0.7 
to $\rm 2.5\times10^{10}\,M_{\odot}$ in the quasar host galaxies. 
The observed CO line widths are spread over a wide range from 
$\rm 160$ to $\rm 860\,km\,s^{-1}$, with a median value of 
about $\rm 360\,km\,s^{-1}$. The gas mass and CO line width distributions of the z$\sim$6 
quasars are consistent with samples of CO-observed SMGs and quasars 
at $\rm 1.4\leq z\leq5$.  

The CO and FIR luminosities of the eight z$\sim$6 quasars 
follow the $\rm L_{FIR}-L'_{CO}$ relationship derived 
 for local spirals, LIRGs, ULIRGs, high-z SMGs, and CO-detected quasars, 
though the weakest CO detection has the largest offset from the trend. 
This is consistent with the idea of co-eval star formation with rapid 
growth of the supermassive black hole in the early quasar-host systems.
The derived SFRs are from $\rm \sim530$ to 
$\rm 2380\,M_{\odot}\,yr^{-1}$. The corresponding star formation 
efficiencies indicated by the ratios of $\rm SFR/M_{gas}$ are consistent with the 
extreme starburst systems at low and high redshifts. 

We investigate the black hole-bulge
correlations of these FIR and CO luminous quasars at z$\sim$6
using the CO measurements. Based on certain assumptions 
of the molecular gas disk size, average inclination 
angle, and $\sigma$-CO line width relation, we 
estimate the bulge dynamical masses and velocity 
dispersions for our sample and compare them to the local 
black hole-bulge relationships. The results suggest 
that the black hole masses of these z$\sim$6 quasars 
are typically an order of magnitude higher than the values expected 
from the present-day relationships, which is consistent 
with the idea that the formation of the SMBHs occurs prior 
to that of the stellar bulges in the massive high-z 
quasar-galaxy systems. However, we also recognize that there 
are large uncertainties in the estimations of $\rm M_{bulge}$ 
and $\rm \sigma$ for individual objects due to unknown gas 
distribution, disk inclination, and dynamics. 
Further high-resolution observations 
of quasar host galaxies should focus on these FIR and 
CO luminous z$\sim$6 quasars to fully understand the black 
hole-bulge evolution at the highest redshift.

\acknowledgments 
This work is based on observations carried out with the IRAM Plateau 
de Bure Interferometer, the Green Bank Telescope (NRAO), the Very Large 
Array (NRAO), and the SHARC-II bolometer camera at the Caltech 
Submillimeter observatory (CSO). IRAM is supported bu INSU/CNRS (France), MPG (Germany) 
and IGN (Spain). The CSO is supported by the NSF under AST-0540882.
The National Radio Astronomy Observatory (NRAO) is a facility of the National
Science Foundation operated under cooperative agreement
by Associated Universities, Inc. 
We thank the anonymous referee for useful comments.
Ran Wang thank Dr. Alexandre Beelen at Institut d'Astrophysique
Spatiale for helpful discussions and suggestions.
We acknowledge support from the Max-Planck Society
and the Alexander von Humboldt Foundation through the
Max-Planck-Forschungspreis 2005. 
Dominik A. Riechers acknowledges support from from NASA through Hubble
Fellowship grant HST-HF-51235.01 awarded by the Space Telescope
Science Institute, which is operated by the Association of
Universities for Research in Astronomy, Inc., for NASA, under contract
NAS 5-26555. Ran Wang acknowledges support of the National Natural Science
Foundation of China grant 10833006 and grant 0816341034.

{\it Facilities:} \facility{IRAM:30m (MAMBO)}, \facility{VLA}, \facility{Sloan (SDSS)} \facility{CSO (SHARC-II)}

\begin{table}
{\scriptsize \caption{Optical and the millimeter measurements of the sample}
\begin{tabular}{lcccccccc}
\hline \noalign{\smallskip}
\hline \noalign{\smallskip}
Name & z & m$\rm _{1450}$ & $\rm M_{1450}$ &  $\rm S_{250GHz}$ & $\rm S_{1.4GHz} $ & Ref. & $\rm L_{bol}$ & $\rm M_{BH}$ \\
     &   &                &      &  mJy          &  $\rm \mu$Jy & & $\rm 10^{13}\,L_{\odot}$  & $\rm 10^{9}\,M_{\odot}$\\
 (1) & (2) & (3) & (4) & (5) & (6) & (7) & (8) & (9) \\
\noalign{\smallskip} \hline \noalign{\smallskip}
J084035.09+562419.9 & 5.85 & 20.04 & -26.66 & 3.20$\pm$0.64 & 12$\pm$9 & [1],[2],[2] & 5.9 & -- \\
J092721.82+200123.7 & 5.79 & 19.87 & -26.78& 4.98$\pm$0.75 & 50$\pm$11& [1],[2],[3]& -- & --  \\
J104433.04$-$012502.2 & 5.80 & 19.21& -27.47 & 1.82$\pm$0.43 & $\rm -15\pm24$ & [4],[3],[5]& 11.8 & 6.4\\
J104845.05+463718.3 & 6.23 & 19.25 & -27.55 & 3.00$\pm$0.40 & $\rm 7\pm13$& [6],[7],[8]& 9.8 & -- \\
J114816.64+525150.3 & 6.43 & 19.03 &-27.82 & 5.00$\pm$0.60 & 55$\pm$12 & [6],[7],[8]& 19.3 & 5.6\\
J133550.81+353315.8 & 5.93 & 19.89 & -26.82 &2.34$\pm$0.50 & $\rm 35\pm10$ & [1],[2],[2]& -- & --\\
J142516.30+325409.0 & 5.85 & 20.62$^{a}$ & -26.09 & 2.27$\pm$0.51 & 20$\pm$20 & [9],[3],[3]& -- & -- \\
J205406.42$-$000514.8 & 6.06 & 20.60 & -26.15 & 2.38$\pm$0.53 & 17$\pm$23 & [10],[3],[3]& -- & -- \\
\noalign{\smallskip} \hline
\end{tabular}\\
Note -- We list the SDSS name in Column (1), the redshifts, AB magenitudes at 
rest-frame 1450$\AA$ from the discovery paper in Column (2), (3), and (4).  
The 250 GHz and 1.4 GHz continuum 
measurements from the literatures are summarized 
and list in Column (5) and (6). 
Column (7) gives the references for the optical, 
millimeter, and radio data, with [1] -Fan et al. (2006a); 
[2]-Wang et al. (2007); [3]-Wang et al. (2008b); 
[4]-Fan et al. (2000); [5]-Petric et al. (2003); [6]-Fan et al. (2003); 
[7]-Bertoldi et al. (2003a); [8]-Carilli et al. (2004); 
[9]-Cool et al. (2006); [10]-Jiang et al. (2008). 
The last two columns list the available measurements of AGN 
bolometric luminosities and black masses from Jiang et al. (2006).
\\
$^{a}$Derived with the absolute AB magnitude at rest-frame 
1450$\rm \AA$ from Cool et al. 2006.
}
\end{table}

\begin{table}
{\scriptsize \caption{Measurements of the CO emission}
\begin{tabular}{lcccccccc}
\hline \noalign{\smallskip}
\hline \noalign{\smallskip} 
Name & transition & $\rm z_{CO}$ &  FWHM & I$\Delta$v & $\rm t_{on,PdBI}$ 
& $\rm s_{con,3mm}$ & $\rm s_{con,350\mu m}$  & $\rm t_{on,SHARC}$ \\
     &            &              &  km s$^{-1}$ & Jy km s$^{-1}$ & hour & mJy & mJy & hour \\
 (1) & (2) & (3) & (4) & (5) & (6) & (7) & (8) & (9) \\
\noalign{\smallskip} \hline \noalign{\smallskip}
\multirow{3}{*}{J0840+5624} & 5$-$4 &\multirow{2}{*}{5.8441$\pm$0.0013}  & \multirow{2}{*}{860$\pm$190} & 0.60$\pm$0.07 & 31.4 & 0.10$\pm$0.05 & \multirow{3}{*}{9.3$\pm$3.2} & \multirow{3}{*}{7.3}\\
                            &6$-$5 &  &    & 0.72$\pm$0.15 & 8.2 & -0.01$\pm$0.10 &  & \\
           & 2$-$1 & -- & -- & $<$0.1$^{a}$ & -- &-- &  & \\
J0927+2001 & 2$-$1 & 5.7722$\pm$0.0006$^{b}$  & --  & $<$0.1& -- &-- &  11.7$\pm$2.4 & 12.0 \\
J1044$-$0125& 6$-$5 & 5.7824$\pm$0.0007 &160$\pm$60  & 0.21$\pm$0.04 & 19.8 & 0.10$\pm$0.05 & $<$13.5 & 2.3 \\
\multirow{2}{*}{J1048+4637} & 6$-$5 & 6.2284$\pm$0.0017 &370$\pm$130 & 0.27$\pm$0.05 & 30.1  
& \multirow{2}{*}{0.13$\pm$0.03} & \multirow{2}{*}{$<$17.4$^{d}$} & \multirow{2}{*}{--}\\
          & 3$-$2 &  --  &  310$^{c}$  & 0.09$\pm$0.02 & -- & -- &  \\
\multirow{2}{*}{J1048NE} & \multirow{2}{*}{6$-$5} & \multirow{2}{*}{6.2259$\pm$0.0019}  & $\sim$550 & 0.23$\pm$0.05 & \multirow{2}{*}{30.1} & \multirow{2}{*}{--} & \multirow{2}{*}{--} & \multirow{2}{*}{--}\\
     &      &  & -- & 0.58$\pm$0.13$^{e}$ & & &  & \\
J1335+3533 & 6$-$5 & 5.9012$\pm$0.0019  & 310$\pm$50  & 0.53$\pm$0.07 & 16.7 & 0.05$\pm$0.05  & $<$13.2 & 3.5 \\
J1425+3254 & 6$-$5 & 5.8918$\pm$0.0018  & 690$\pm$180 & 0.59$\pm$0.11 & 16.1 & 0.09$\pm$0.04  & $<$7.8 & 4.7 \\
J2054$-$0005& 6$-$5 & 6.0379$\pm$0.0022 & 360$\pm$110 & 0.34$\pm$0.07 & 16.6 & 0.10$\pm$0.05  & -- & --\\
\noalign{\smallskip} \hline
\end{tabular}\\
Note -- Column (1), source names; Column (2), CO 
transitions observed in this work; Column (3), Column (4), 
and Column (5), redshifts, line widths, and flux measured with 
the CO detection; Column (6), on-source time; Column (7), 
dust continuum measurements under the CO line spectrum observed with PdBI; 
Column (8) and (9), dust continuum measurements and on-source 
integration time with SHARC-II at 350 $\mu$m.\\
$^{a}$Upper limits derived with the channel-to-channel rms noise. One should be cautious
with this value as there is a small-scale structure close to
the line frequency which preclude
detection of weak line emission with peak
flux density of a few hundred $\rm \mu Jy$ (Wagg et al. 2008);
$^{b}$Carilli et al. (2007), measured with
the CO (6-5) and (5-4) line emission;
$^{c}$line width calculated with the 50 MHz channel width;
$^{d}$Wang et al. (2008a); 
$^{e}$Primary beam attenuation-corrected 
line flux for J1048NE (see Figure 1).
}
\end{table}

\begin{table}
{\scriptsize 
\caption{Redshift measurements}
\begin{tabular}{lccccccccc}
\hline \noalign{\smallskip} \hline \noalign{\smallskip}
  Name &  z$_{CO}$ & z$\rm _{Ly\alpha}$ & Ref & z$\rm _{MgII}$ & Ref & z$\rm _{CVI}$ & Ref & z$\rm _{UV}^{c}$ &  Ref \\
  (1)  & (2)       &   (3)          & (4) &  (5)       & (6) & (7)       & (8) & (9)     &  (10) \\
\hline\noalign{\smallskip} \hline \noalign{\smallskip}
J0840+5624   & 5.8441$\pm$0.0013 & 5.85$\pm$0.02   & (1) &  & & 5.774 & (2) & & \\
J0927+2001   & 5.7722$\pm$0.0006 & 5.79$\pm$0.02 & (1) & & & & & & \\
J1044$-$0125 & 5.7824$\pm$0.0007 & &  & 5.78 & (3) & 5.745$\pm$0.030 & (4) &  5.80$\pm$0.02 & (5) \\
             &                   & &  &      &                       &                 &   & 5.778$\pm$0.005& (12) \\
J1048+4636   & 6.2284$\pm$0.0017 &6.23$\pm$0.05 &  (6) & 6.203 &  (7) & & & & \\
             &                   &              &                  & 6.22 & (8) & & & & \\
             &                   &              &                  & 6.193  & (9) & & & & \\
J1148+5251   & 6.4192$\pm$0.0009$^{a}$& 6.43$\pm$0.05& (6) & 6.41$\pm$0.01& (10) &  & & 6.421 & (2) \\
             & 6.4189$\pm$0.0006$^{b}$&              &     & 6.40         & (8) &  &  &      &        \\
J1335+3533   & 5.9012$\pm$0.0019 &5.93$\pm$0.04 & (1) & & & & & & \\
J1425+3254   & 5.8918$\pm$0.0018 &5.85$\pm$0.02 & (11) & & & & & & \\
J2054$-$0005 & 6.0379$\pm$0.0022 & 6.062$\pm$0.004 & (13) & & & & & & \\
\noalign{\smallskip} \hline
\end{tabular}\\
$^{a}$Redshift measured from the CO (7-6) line (Bertoldi et al. 2003b).
$^{b}$Redshift measured from the CO (6-5) line (Bertoldi et al. 2003b).
$^{c}$Redshift measured from other UV emission lines, such as the 
the C{\tiny III]} $\rm \lambda 1909\AA$, O{\tiny I}+Si{\tiny II} $\rm \lambda 1302\AA$, 
and Si{\tiny IV}+O{\tiny IV} $\rm\lambda 1400\AA$.\\
References: (1) Fan et al. 2006a; (2) Ryan-Weber et al. 2009; (3) Freudling et al. 2003; 
(4) Goodrich et al. 2001; (5) Fan et al. 2000; (6) Fan et al. 2003; (7) Iwamuro et al. 2004; 
(8) Maiolino et al. 2004a; (9) Maiolino et al. 2004b; (10) Willott et al. 2003; (11) Cool et al. 2006; 
(12) Jiang et al. 2007; (13) Jiang et al. 2008.
}
\end{table} 

\begin{table}
{\scriptsize 
\caption{Derived parameters}
\begin{tabular}{lcccccc}
\hline \noalign{\smallskip} \hline \noalign{\smallskip}
Name & $\rm {L'_{CO}}$ & $\rm L'_{CO(1-0)}$ & $\rm L_{FIR}$ & SFR  & $\rm M_{dyn}sin^2i$ & $\rm M_{gas}$ \\
& $\rm 10^{10}\,K\,km\,s^{-1}\,pc^2$ & $\rm 10^{10}\,K\,km\,s^{-1}\,pc^2$ 
& $\rm 10^{12}\,L_{\odot}$& $\rm M_{\odot}\,yr^{-1}$ & $\rm 10^{10}\,M_{\odot}$ & $\rm 10^{10}\,M_{\odot}$\\
(1) & (2) & (3) &(4) & (5) & (6) & (7) \\
\hline\noalign{\smallskip} \hline \noalign{\smallskip}
J0840+5624  & 2.8$\pm$0.3$^{a}$& 3.2$\pm$0.4 & 7.2 (5.7) & 1460  &24.2/4.2 & 2.5 \\
J0927+2001  & 2.0$\pm$0.3$^{a}$& 2.3$\pm$0.4 & 8.5 (6.7) & 1740  &11.8 & 1.8 \\
J1044$-$0125& 0.7$\pm$0.1$^{b}$& 0.8$\pm$0.2 & 5.2 (2.0) & 530  &0.8  & 0.7 \\
J1048+4637  & 1.0$\pm$0.2$^{b}$&1.2$\pm$0.2  &5.9 (2.5) & 650  &4.5  & 1.0 \\ 
J1148+5251  & 3.0$\pm$0.3$^{c}$&3.0$\pm$0.3  & 13.6 (9.2) & 2380  &4.6  & 2.4 \\
J1335+3533  & 1.7$\pm$0.2$^{b}$&2.2$\pm$0.3  &5.5 (3.8) & 970  &3.1  & 1.8 \\
J1425+3254  & 1.9$\pm$0.4$^{b}$&2.5$\pm$0.5  &5.4 (4.5) & 1160 &15.6 & 2.0 \\
J2054$-$0005& 1.2$\pm$0.2$^{b}$&1.5$\pm$0.3  &5.5 (4.6) & 1180  &4.2  & 1.2 \\
\noalign{\smallskip} \hline
\end{tabular}\\
Note -- Column (1), source name; Column (2), line luminosities 
of a. the CO (5-4) transition (Carilli et al. 2007; this work), 
b. the CO (6-5) transition (this work), and c. the CO (3-2) 
transition (Walter et al. 2003; Riechers et al. 2009) 
in $\rm 10^{10}\,K\,km\,s^{-1}\,pc^2$ (see e.g., SV05 for the calculation), which are used to
derive $\rm L'_{CO(1-0)}$ in the next column (See \S 4.1 for details); 
Column (3), the derived CO (1-0) luminosity; Column (4), FIR luminosity 
derived with the submillimeter and millimeter continuum data 
from the literature, and the values quoted in brackets are the AGN-corrected 
FIR luminosity (See \S 4.4 for details); Column (5) star formation 
rate derived with the AGN-corrected FIR luminosity; Column (6) dynamical 
masses derived with the observed CO line widths as was described 
in \S 5.2.1. The two values for J0840+5624 are derived with the single-Gaussian fitted
FWHM of $\rm 860\,km\,s^{-1}$ and the average peak offset
of $\rm 270\,km\,s^{-1}$, respectively. We adopt the CO line widths of $\rm FWHM=600\,km\,s^{-1}$
for J0927+2001 from Carilli et al. (2007), and $\rm FWHM=297\,km\,s^{-1}$
from Walter et al. (2009) for J1148+5251. Column (7), 
molecular gas masses derived with $\rm L'_{CO(1-0)}$. 
}
\end{table}

\begin{figure}[h]
\includegraphics[height=4.8in,width=3.0in,angle=-90]{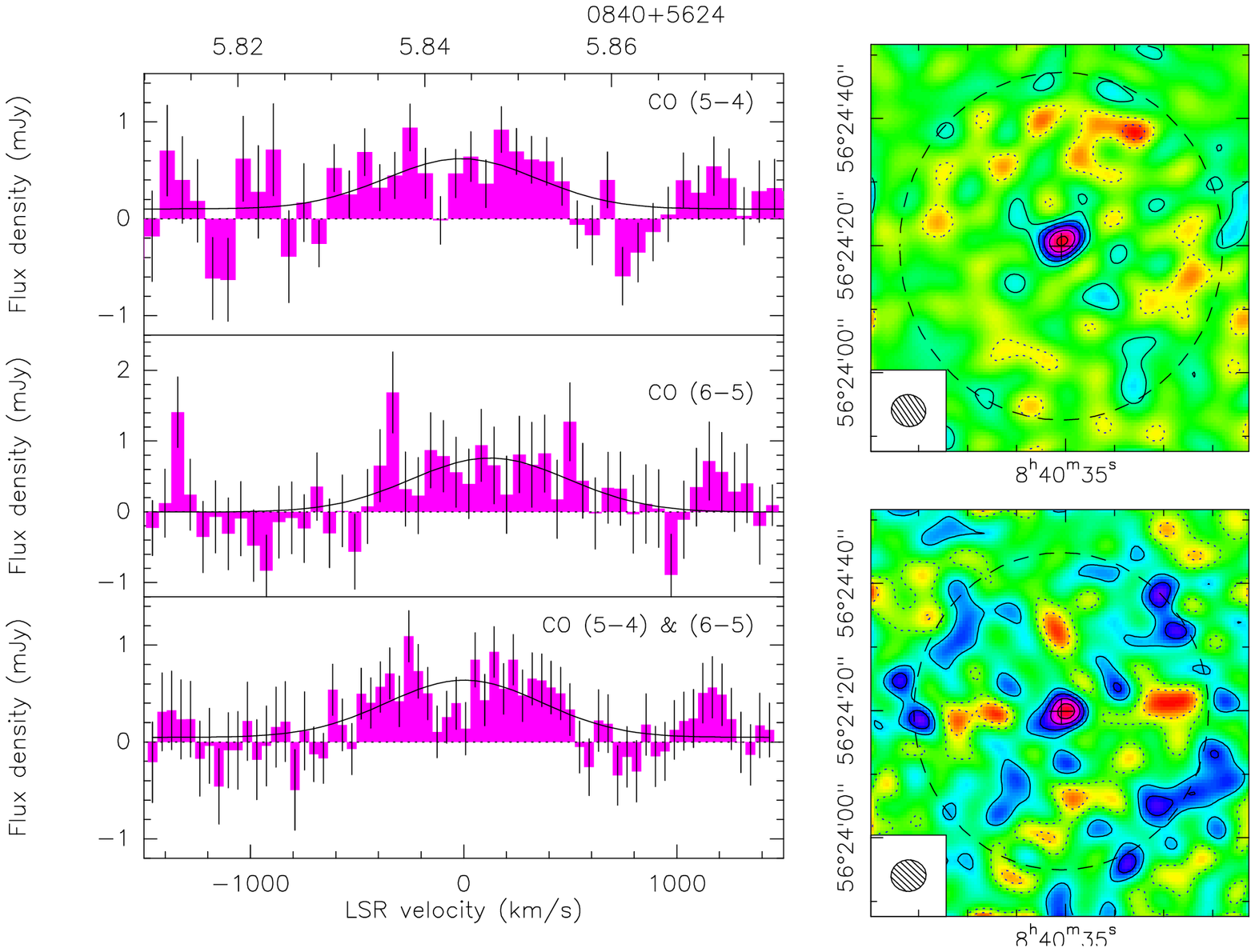}\\
\includegraphics[height=4.8in,width=1.6in,angle=-90]{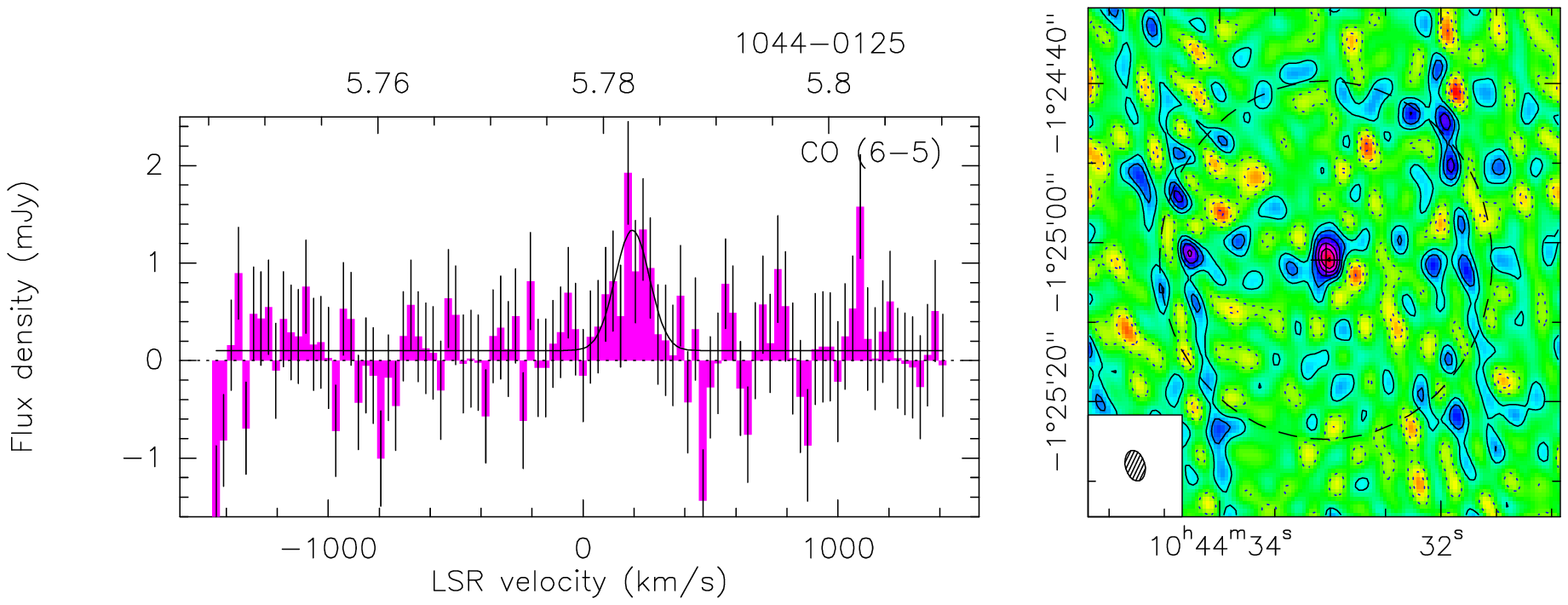}
\caption{The CO line spectra (left) and velocity-integrated 
images (right) of the six CO-detected quasars at z$\sim$6. 
The top abscissa of the left plots give the redshift range of 
the spectra windows, and the zero velocities correspond to the 
redshifts of z=5.844, 5.778, 6.200, 5.950, 5.877, 
and 6.070. The error bars in the spectra denote the 1$\rm \sigma$ 
rms noise in each channel, and 
the solid lines show Gaussian fits to the line emission. 
The third spectrum for J0840+5624 is a combined spectrum 
obtained by merging the data of the two detected CO transitions. 
In the spectrum of the NE source close to J1048+4637,  
the gray line indicates the primary beam attenuation-corrected line
profile, assuming a line width of 370 km s$^{-1}$.
The crosses on the images indicate the optical quasar position, and the dashed 
circles shows the primary beam of the PdBI. 
On the map of J1044-0125, the spurious
pattern to the east and west of the source is caused by the substantial
sidelobes due to the strong signal from the central source.
}
\end{figure} 

\begin{figure}[h]
\figurenum{1}
\includegraphics[height=3.3in,width=4.8in]{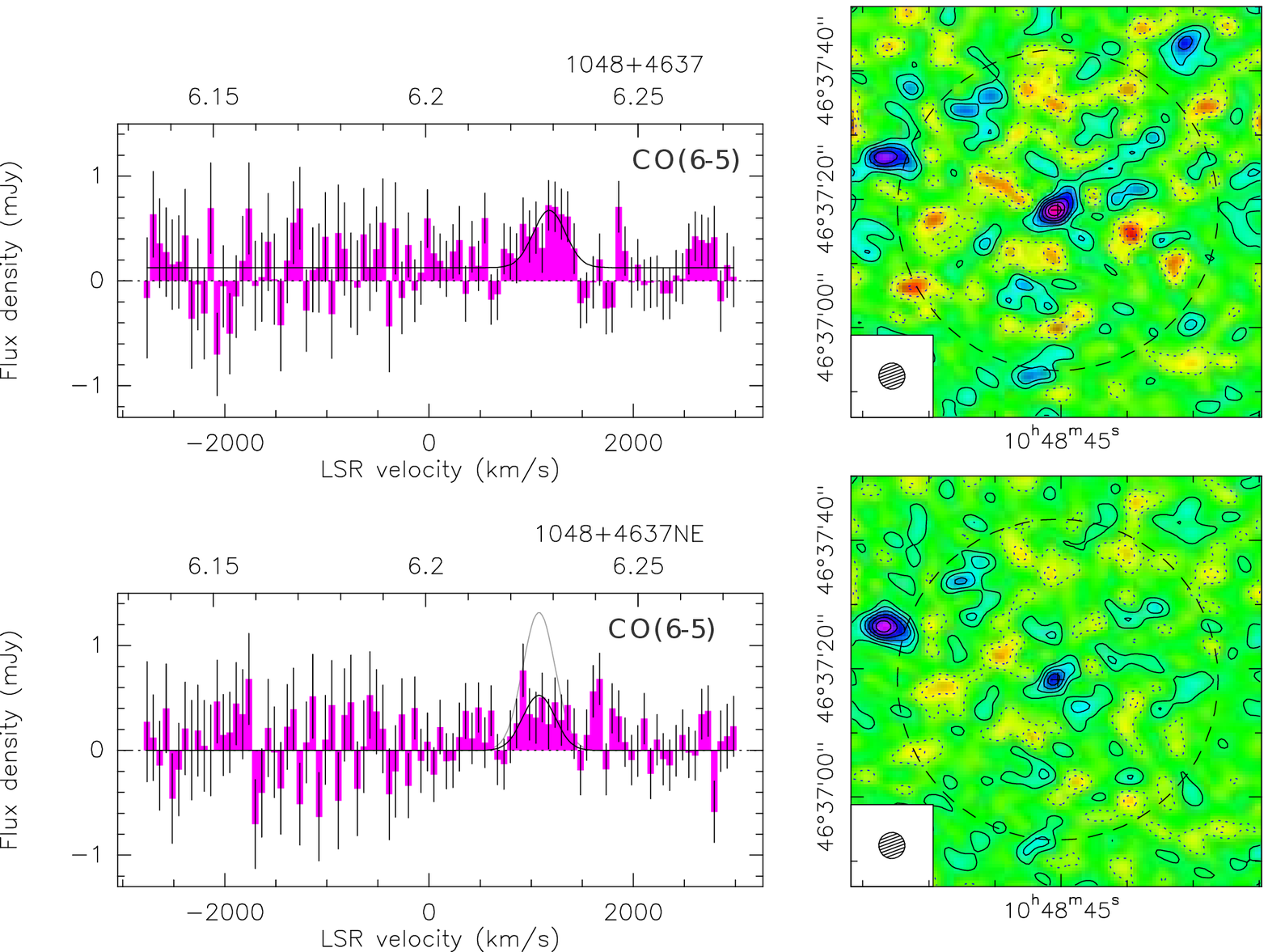}\\
\includegraphics[height=4.8in,width=1.6in,angle=-90]{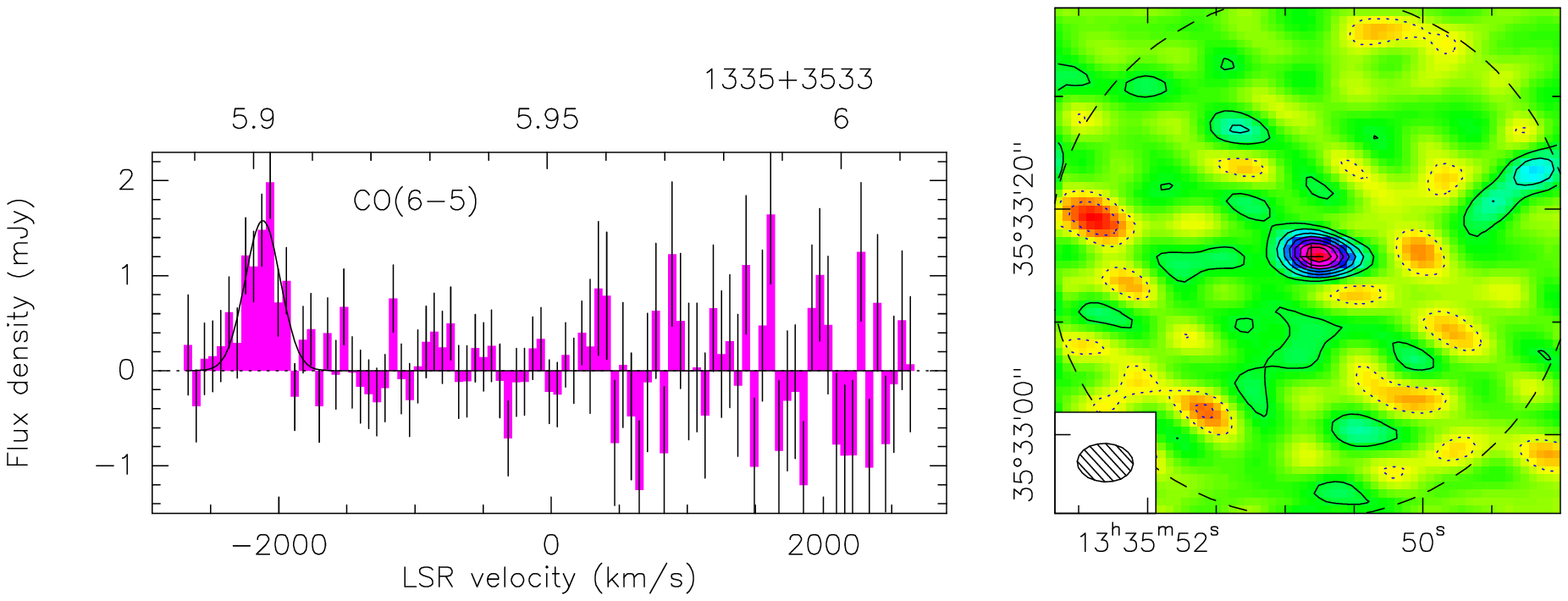}\\
\includegraphics[height=4.8in,width=1.6in,angle=-90]{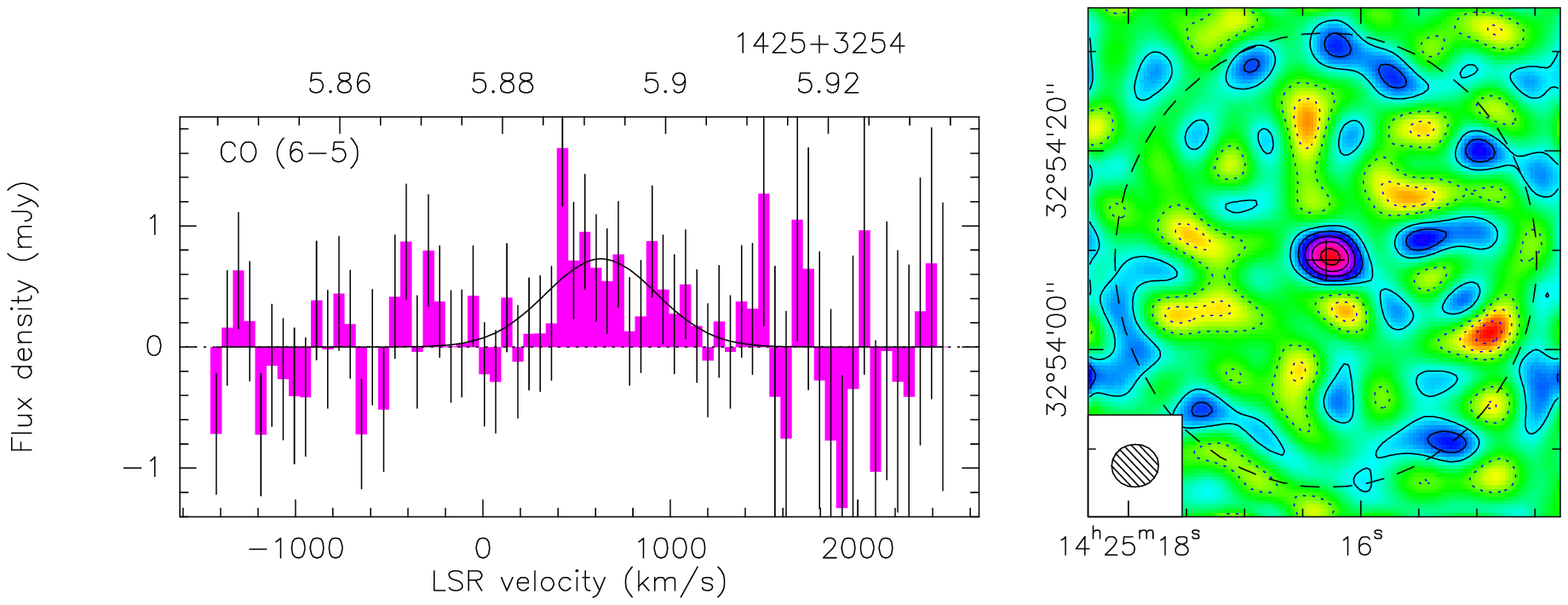}\\
\includegraphics[height=4.8in,width=1.6in,angle=-90]{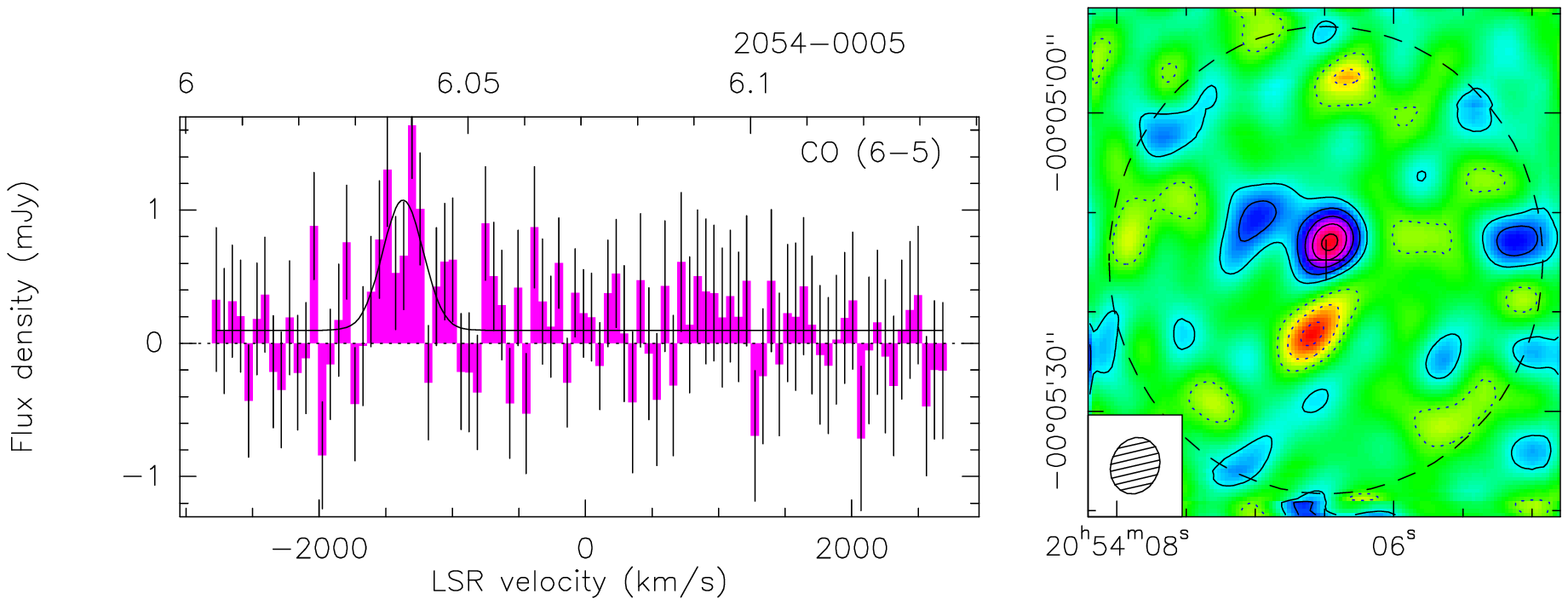}
\caption{Continued.}
\end{figure} 

\begin{figure}
\epsscale{1.3}
\plottwo{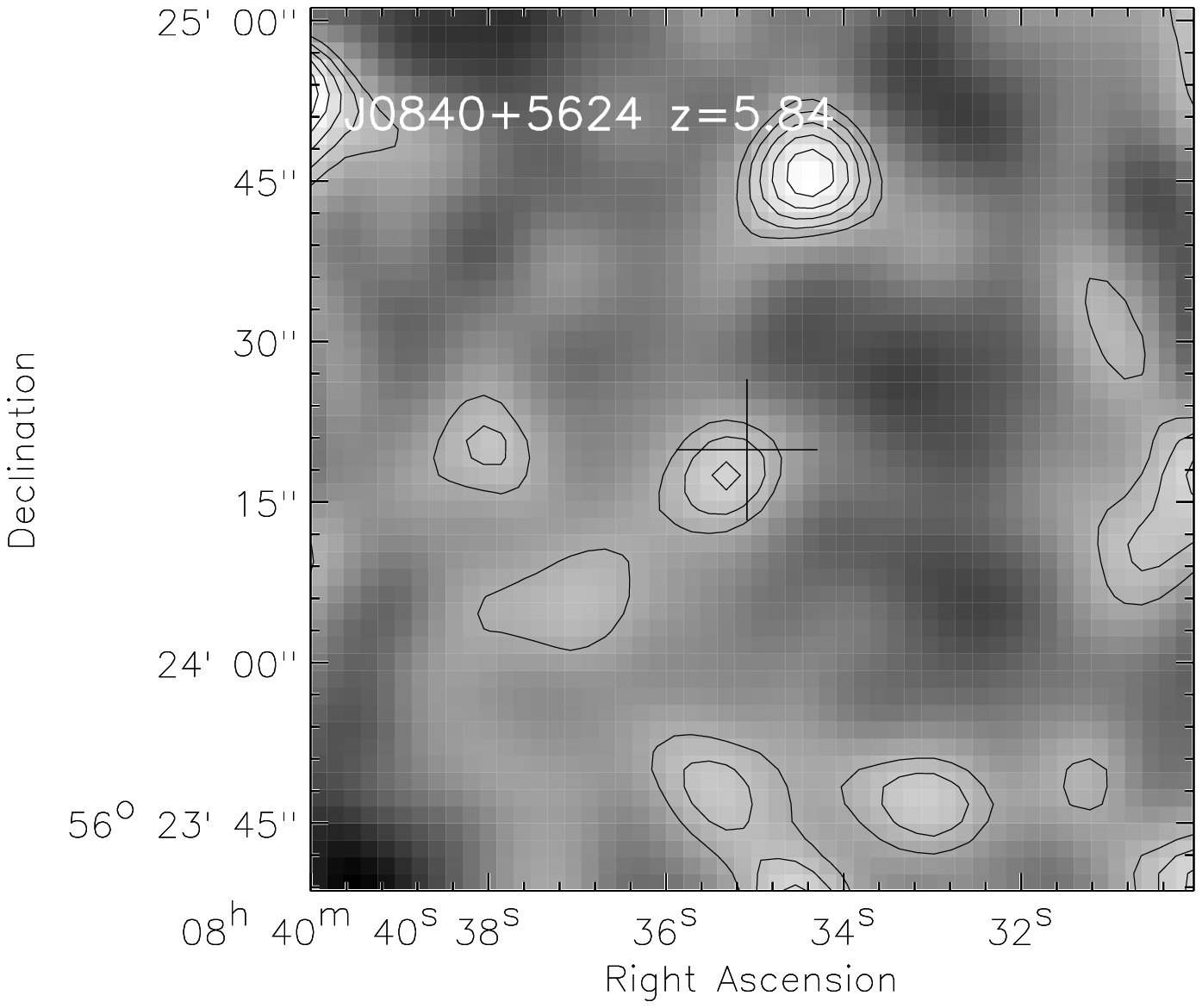}{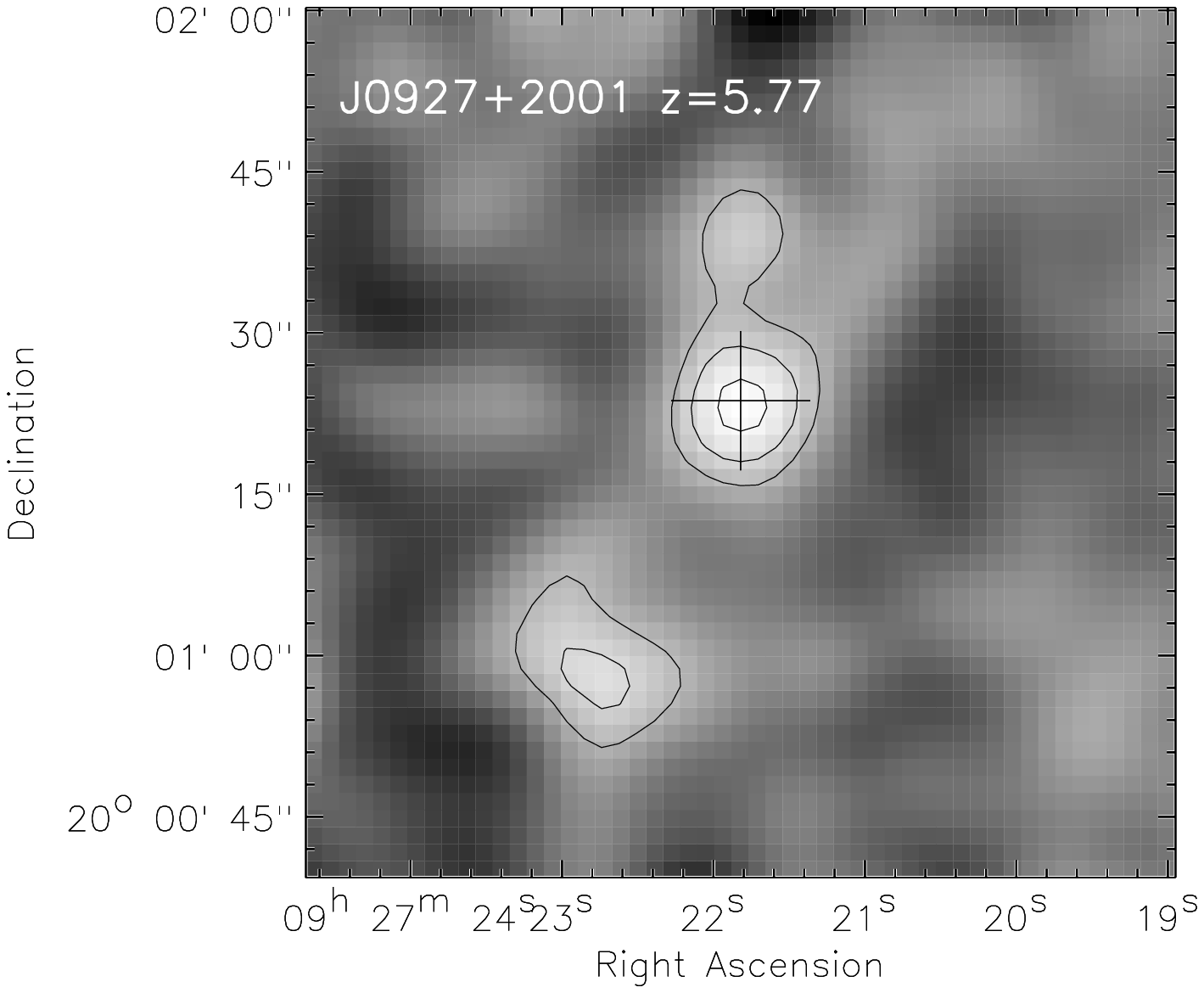}
\caption{The 350 $\mu$m maps of the SHARC-II marginally detected 
source J0840+5624 (left) and SHARC-II detected source J09227+2001 (right), 
smoothing to a beam size of $\rm FWHM=12.4''$. The contour levels 
are (2,3,4,5,6)$\rm \times2.2\,mJy\,beam^{-1}$ for J0840+5624, 
and (2,3,4)$\rm \times2.7\,mJy\,beam^{-1}$ for J0927+2001. 
The crosses denote the optical quasar positions.}
\end{figure}

\begin{figure}
\epsscale{1.0}
\plottwo{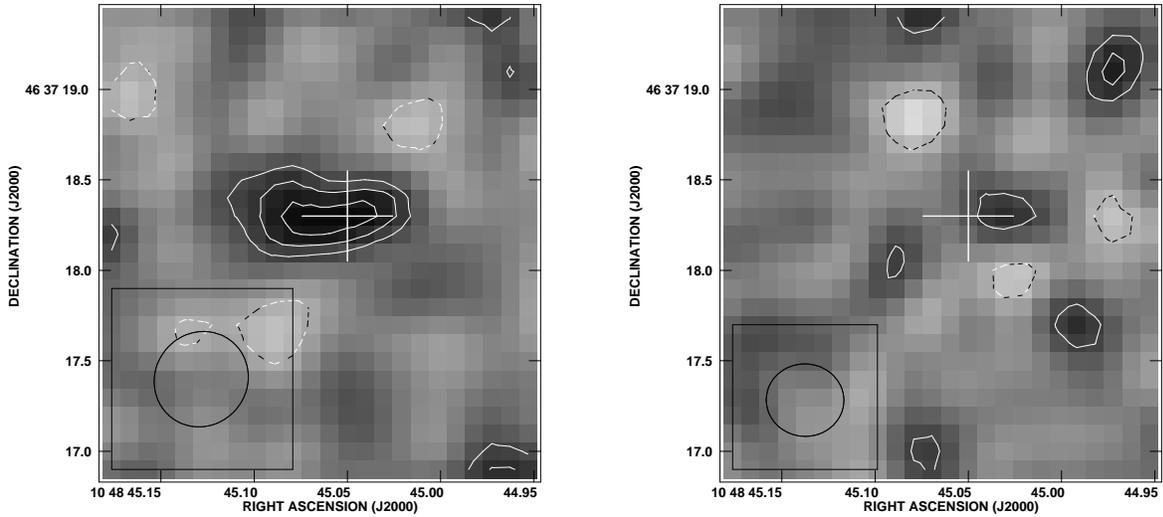}{fig3b.ps}
\caption{Left--The VLA Q band image of the CO (3-2) line 
emission from J1048+4637 detected 
at 47.865 GHz (Left) and a 48 GHz image averaged 
over the line-free channels (right). 
The 1$\rm \sigma$ rms values are 70 $\mu$Jy beam$^{-1}$ 
and 65 $\mu$Jy beam$^{-1}$ for the left and right images, respectively.
The contour levels are (-2, 2, 3, 4) $\times$ 60 $\mu$Jy beam$^{-1}$ 
for both images. The ellipses  
indicate the beam sizes of $\rm 0.5''\times0.5''$ (FWHM) 
for the left plot and $\rm 0.4''\times0.4''$ for the right one, 
and the cross marks the optical quasar position.}
\end{figure}

\begin{figure}
\epsscale{1.15}
\plottwo{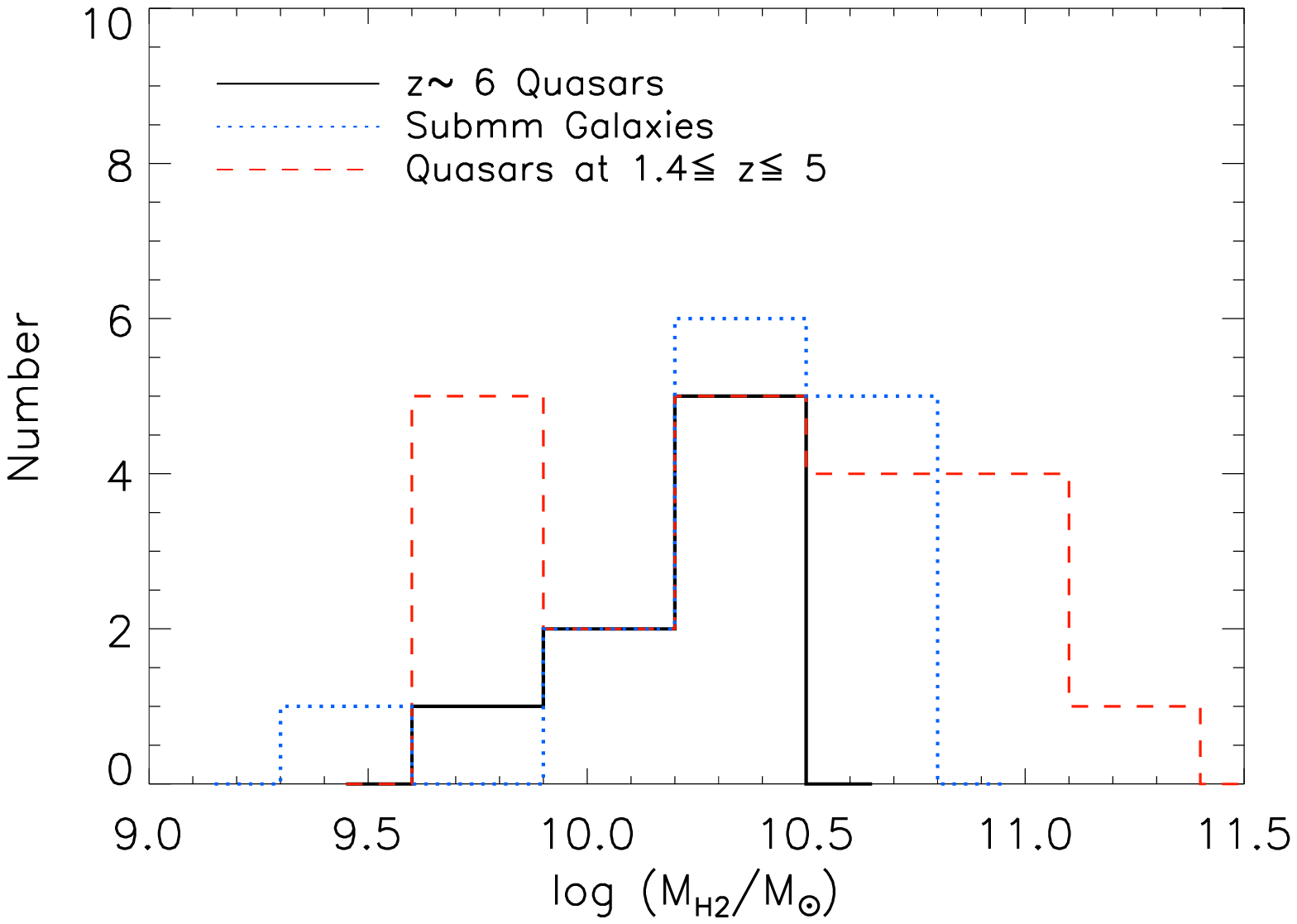}{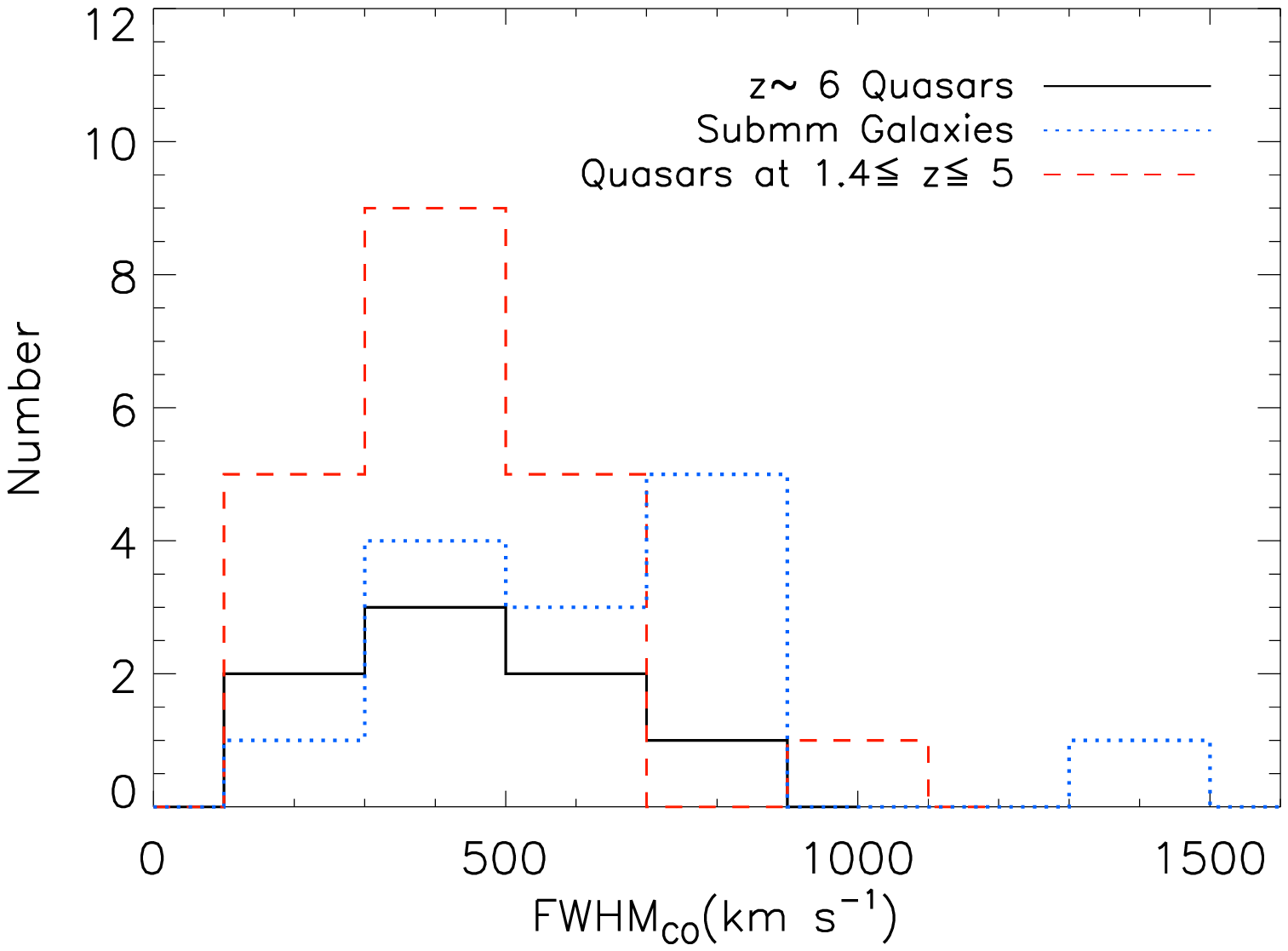}
\caption{The molecular gas mass (left) and 
CO line width (right) distributions of the 
CO-detected z$\sim$6 quasars, the SMGs, and 
$\rm 1.4\leq z\leq5$ quasars. }
\end{figure}

\begin{figure}
  \epsscale{0.9}
  \plotone{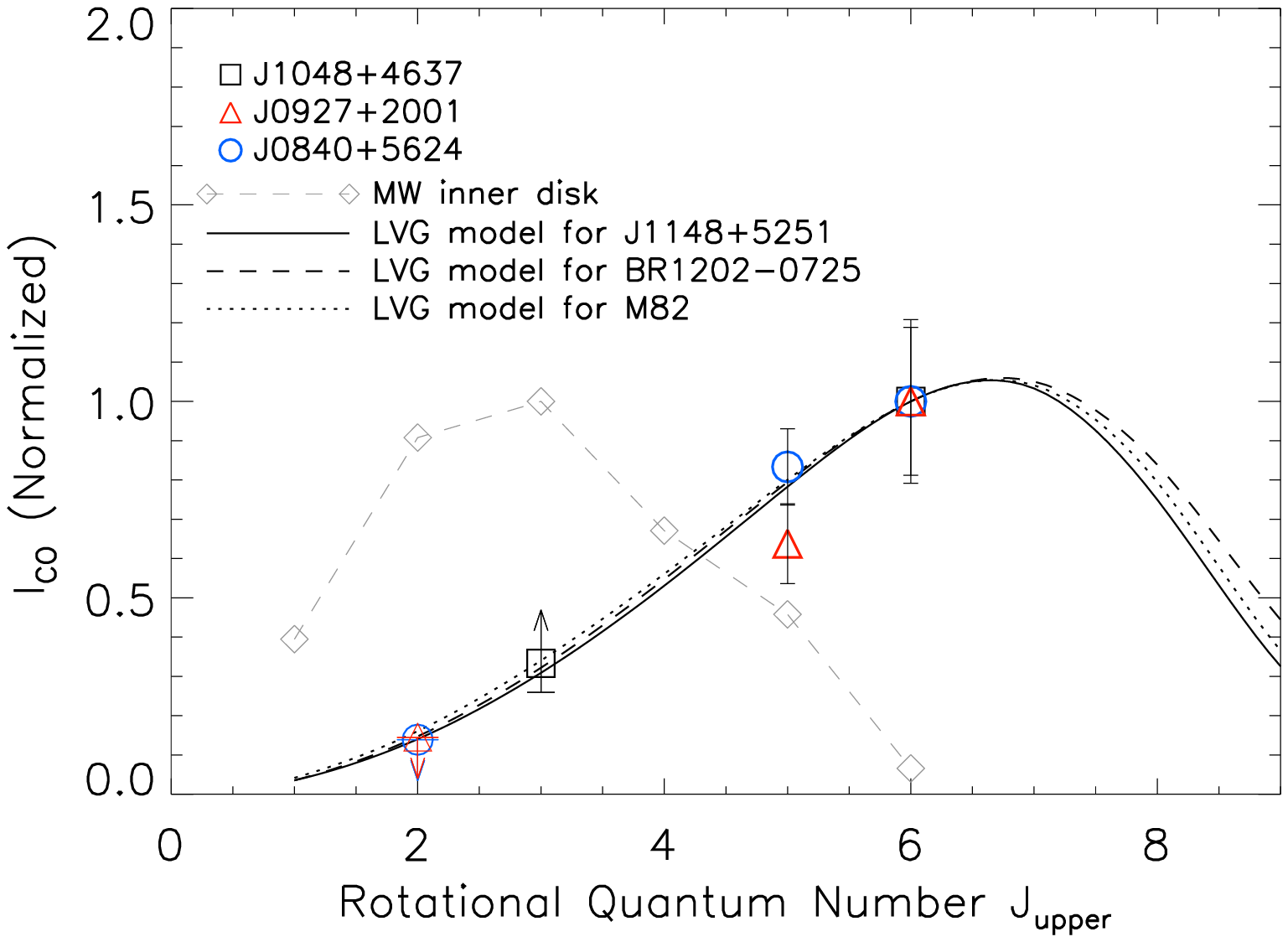}
  \caption{CO spectral energy distributions of J0840+5624 (blue circles), 
  J0927+2001 (red triangles), and J1048+4637 (black squares). 
  The arrow with black square represents the VLA 
  detection of the CO (3-2) line emission from J1048+4637 in 
  a 50 MHz channel, which provides a lower limit of the total  
  line intensity due to the narrow bandwidth.
  The error bars 
  show the 1$\sigma$ uncertainties of the integrated line 
  intensities. 
  The solid line and dashed line show the LVG models   
  of J1148+5251 with $\rm T_{kin}=50\,K$, $\rm \rho_{gas}(H_2)=10^{4.2}\,cm^{-3}$ 
  from Riechers et al. (2009) 
  and BR 1202-0725 with $\rm T_{kin}=60\,K$, $\rho_{gas}(H_2)=10^{4.1}\,cm^{-3}$ 
  from Riechers et al. (2006), respectively. 
  The dotted line represent the LVG model of the high excitation 
component in the nearby starburst galaxy M82 
with $\rm T_{kin}=50\,K$, $\rm \rho_{gas}(H_2)=10^{4.2}\,cm^{-3}$ from Wei$\ss$ et al. (2005b)
  We also plotted the CO excitation ladder of the Milky Way inner 
  disk region for comparison (diamonds with light dashed line, Fixsen et al. 1999).
  The CO line intensities are normalized to the CO (3-2) line for 
the Milky Way, and CO (6-5) for other sources and models.
}
\end{figure}

\begin{figure} 
\epsscale{0.9}
\plotone{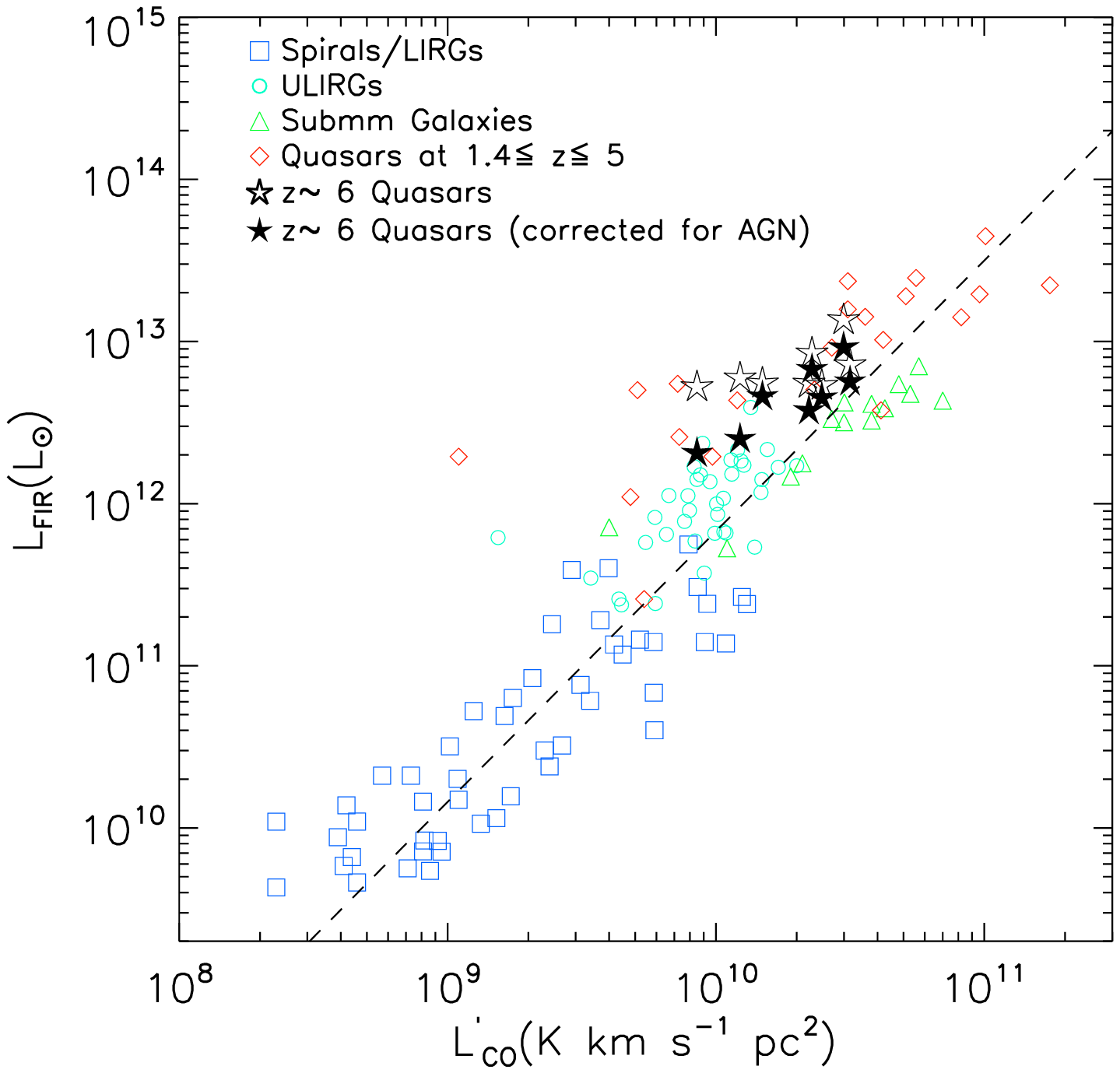}
\caption{ $\rm L_{FIR}$ vs. CO luminosity ($\rm L'_{CO}$). 
The data of samples at z$\leq$5 are taken from Gao \& Solomon (2004), Greve et al. (2005), 
Tacconi et al. (2006), Solomon et al. (1997), SV05, 
Riechers et al. (2006), Maiolino et al. (2007) and Coppin et al. (2008a). 
The dashed line represents the relationship $\rm L_{FIR}\propto {L'_{CO}}^{1.67}$ 
fitted to the samples of local spirals, LIRGs, ULIRGs and SMGs. 
The open stars represent the eight CO detected quasars at z$\sim$6 
with $\rm L_{FIR}$ estimated from (sub)mm observations, and the filled 
stars denote the values when contributions from AGN are subtracted (see \S 4.4 for details).}
\end{figure}


\begin{figure} 
\epsscale{0.9}
\plotone{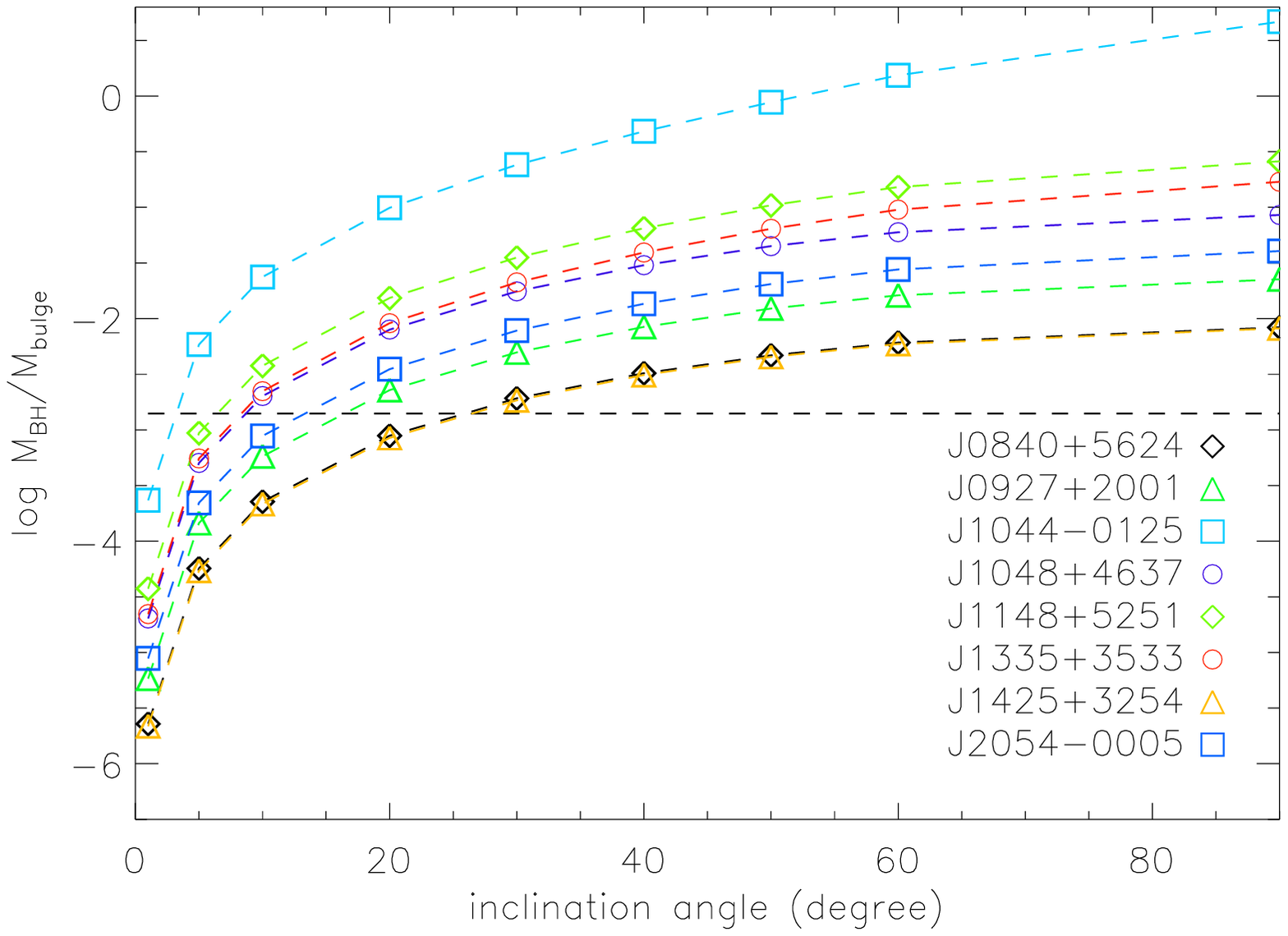}
\caption{The derived ratios between the black hole and the 
bulge stellar masses versus the inclination angles of the 
molecular disk. The bulge dynamical masses ($\rm M_{dyn}\,sin^2{\it i}$) 
are derived using the CO FWHMs from single-Gaussian spectral 
fitting, and the bulge stellar masses ($\rm M_{bulge}$) 
are estimated with $\rm M_{dyn}-M_{gas}$. The plot shows 
how $\rm M_{BH}/M_{bulge}$ compares to present day value of 0.0014 (dashed line) 
with different assumptions of inclination angles.}
\end{figure}

\begin{figure}
\epsscale{0.9}
\plotone{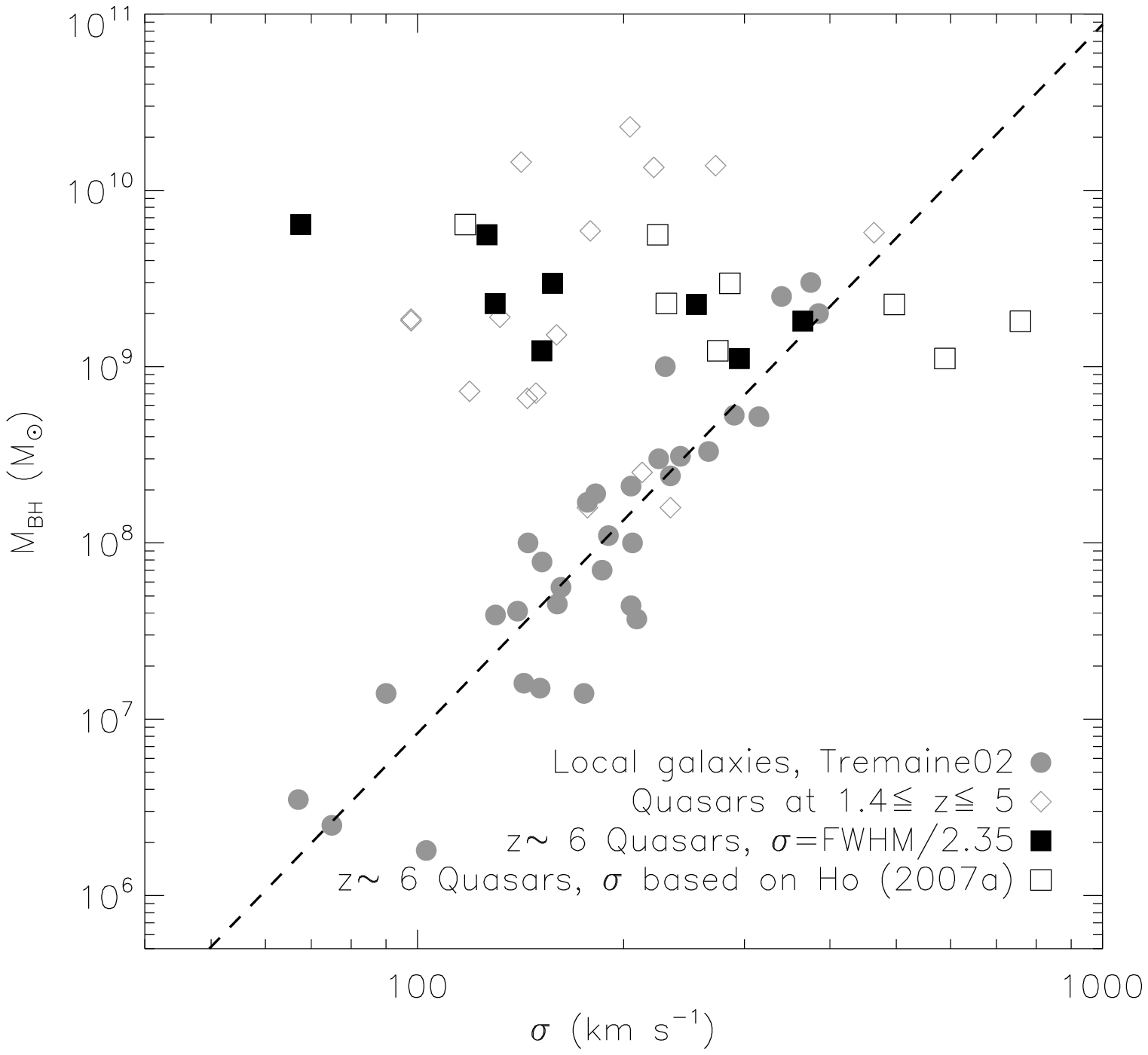}
\caption{The black hole masses of the z$\sim$6 quasars 
versus their bulge velocity dispersion ($\sigma$). 
The dashed line denotes the local $\rm M_{BH}-\sigma$ 
relationship of 
$\rm log\,(M_{BH}/M_{\odot})=8.13+4.02log(\sigma/200\,km\,s^{-1})$ (Tremaine et al. 2002).
The filled circles represent the local galaxies from Tremaine et al. (2002). 
The filled squares and open diamonds 
are for the z$\sim$6 and $\rm 1.4\leq z\leq5$ quasar samples, 
respectively, with $\sigma$ derived from the observed CO 
line width using $\sigma\approx FWHM/2.35$. 
The open squares show the $\sigma$ values derived 
with the method described in Ho (2007a) assuming 
an average inclination angle of $\rm 40\degrees$ 
for the z$\sim$6 quasars. 
}
\end{figure}


\begin{thebibliography}{}
\bibitem[]{} Beelen, A., Cox, P., Benford, D. J., Dowell, C. D., Kov$\rm \acute{a}$cs, A., Bertoldi, F., Omont, A., \& Carilli, C. L., 2006, ApJ, 642, 694
\bibitem[]{} Benford, D. J., Cox, P., Omont, A., Phillips, T. G., \& McMahon, R. G. 1999, ApJ, 518, L65
\bibitem[]{} Bertoldi, F., Carilli, C. L., Cox, P., Fan, X., Strauss, M. A., Beelen, A., Omont, A., \& Zylka, R., 2003, A\&A, 406, L55
\bibitem[]{} Bertoldi, F. et al. 2003b, A\&A, 409, L47
\bibitem[]{} Brown, R. L. \& Vanden Bout, P. A. 1992, ApJ, 397, L19
\bibitem[]{} Carilli, C. L. et al. 2002, AJ, 123, 1838
\bibitem[]{} Carilli, C. L. et al. 2004, AJ, 128, 997
\bibitem[]{} Carilli, C. L. \& Wang, R. 2006, AJ, 131, 2763
\bibitem[]{} Carilli, C. L. et al. 2007, ApJ, 666, L9
\bibitem[]{} Chapman, S. C., Blain, A. W., Smail, I., \& Ivison, R. J., ApJ, 622, 772
\bibitem[]{} Cool, R. J. et al., 2006, AJ, 132, 823
\bibitem[]{} Coppin, K. E. K. et al. 2008a, MNRAS, 389, 45
\bibitem[]{} Coppin, K. E. K. et al. 2008b, MNRAS, 384, 1597
\bibitem[]{} Cox, P. et al. 2002, A\&A, 387, 406
\bibitem[]{} Daddi, E., Dannerbauer, H., Elbaz, D., 
Dickinson, M., Morrison, G., Stern, D. \& Ravindranath, S. 2008, ApJ, 673, L21
\bibitem[]{} Daddi, E. et al. 2009, ApJ, submitted, astro-ph/0911.2776
\bibitem[]{} Downes, D., \& Solomon, P. M. 1998, ApJ, 507, 615
\bibitem[]{} Elitzur, M. 2008, NewAR, 52, 274
\bibitem[]{} Elvis, M. et al. 1994, ApJS, 95, 1
\bibitem[]{} Fan, X. et al. 2000, AJ, 120, 1167
\bibitem[]{} Fan, X. et al. 2003, AJ, 125, 1649
\bibitem[]{} Fan, X. et al. 2006a, AJ, 131, 1203
\bibitem[]{} Fan, X. et al. 2006b, AJ, 132, 117
\bibitem[]{} Fan, X., Carilli, C. L., \& Keating, B. 2006c, ARA\& A, 44, 415
\bibitem[]{} Fixsen, D. J., Bennett, C. L., \& Mather, J. C. 1999, ApJ, 526, 207
\bibitem[]{} Freudling, W., Corbin, M. R., Korista, K. T. 2003, ApJ, 587, L67
\bibitem[]{} Gao, Y., \& Solomon, P. M. 2004, ApJS, 152, 63
\bibitem[]{} Goodrich, R. W., et al. 2001, ApJ, 561, L23
\bibitem[]{} Greve, T. R. et al. 2005, MNRAS, 359, 1165
\bibitem[]{} G\"usten, R., Philipp, S. D., Wei$\ss$, A., Klein, B. 2006, A\&A, 454, L115
\bibitem[]{} Guilloteau, S., \& Lucas, R. 2000, ASPC, 217, 299
\bibitem[]{} Ho, L. C. 2007a, ApJ, 669, 821
\bibitem[]{} Ho, L. C. 2007b, ApJ, 668, 94
\bibitem[]{} Hopkins, P. F., Hernquist, L., Cox, T. J., Robertson, B., \& Krause, E. 2007a,
ApJ, 669, 67
\bibitem[]{}Hopkins, P. F., Hernquist, L., Cox, T. J., Robertson, B., \& Krause,
E. 2007b, ApJ, 669, 45
\bibitem[]{} Iwamuro, F,. Kimura, M., Eto, S., Maihara, T., Motohara,
K., Yoshii, Y., \& Doi, M. 2004, ApJ, 614, 69
\bibitem[]{} Isobe, T, Feigelson, E. D., Akritas, M. G. \& Babu, G. J. 1990, ApJ, 364, 104
\bibitem[]{} Jiang, L. et al. 2006, AJ, 132, 2127
\bibitem[]{} Jiang, L. et al. 2007, AJ, 134, 1150
\bibitem[]{} Jiang, L. et al. 2008, AJ, 2008, 135, 1057
\bibitem[]{} Kauffmann, G. \& Haehnelt, M. 2000, MNRAS, 311, 576
\bibitem[]{} Kennicutt, R. C. 1998, ARA\&A, 36, 189
\bibitem[]{} Kreysa, E. et al. 1998, SPIE, 3357, 319
\bibitem[]{} Kov$\rm \acute a$cs, A. 2006a, PhD thesis, Caltech
\bibitem[]{} Kov$\rm \acute{a}$cs, A., Chapman, S. C., Dowell, C. D., Blain, A. W., Ivison, R. J., Smail, I., \& Phillips, T. G., 2006b, ApJ, 650, 592
\bibitem[]{} Maiolino, R., Oliva, E., Ghinassi, F., Pedani, M., Mannucci, F., 
Mujica, R., \& Juarez, Y. 2004a, A\&A, 420, 889
\bibitem[]{} Maiolino, R., Schneider, R., Oliva, E., Bianchi, S.,
Ferrara, A., Mannucci, F., Pedani, M., Roca Sogorb, M. 2004b, Nature, 431, 533
\bibitem[]{} Maiolino, R. et al. 2007, A\&A, 472, L33
\bibitem[]{} Marconi, A., \& Hunt, L. K. 2003, ApJ, 589, L21
\bibitem[]{} Narayanan, D. et al. 2008, ApJS, 174, 13
\bibitem[]{} Nelson, C. H. 2000, ApJ, 544, L91
\bibitem[]{} Neri, R. et al. 2003, ApJ, 597, L113
\bibitem[]{} Omont, A., Petitjean, P. Guilloteau, S. et al. 1996a, Nature, 382, 428
\bibitem[]{} Omont, A., McMahon, R. G., Cox, P., Kreysa, E., Bergeron,
J., Pajot, F., \& Storrie-Lombardi, L. J. 1996b, A\&A, 315, 1
\bibitem[]{} Omont, A., Beelen, A., Bertoldi, F., McMahon, R. G.,
Carilli, C. L., \& Isaak, K. G. 2003,A\& A, 398, 857
\bibitem[]{} Peng, C. Y., Impey, C. D., Ho, L. C., Barton, E. J., \& Rix, H.-W. 2006a, ApJ,
640, 114
\bibitem[]{} Peng, C. Y., Impey, C. D., Rix, H.-W., Kochaneck, C. S., Keeton, C. R., Falco,
E. E., Leha´r, J., \& McLeod, B. A. 2006b, ApJ, 649, 616 
\bibitem[]{} Petric, A. O., Carilli, C. L., Bertoldi, F., Fan, X., Cox, P., Strauss, M. A., Omont, A., \& Schneider, D. P. 2003, AJ, 126, 15
\bibitem[]{} Priddey, R. S., Isaak, K. G., McMahon, R. G., Roboson, E. I., \& Pearson, C. P. 2003, MNRAS, 344, L74
\bibitem[]{} Richards, G. T., Vanden Berk, D. E., Reichard, T. A.,
Hall, P. B., Schneider, D. P., SubbaRao, M., Thakar, A. R., \& York, D. G. 2002, AJ, 124, 1
\bibitem[]{} Riechers, D. A. et al. 2006, ApJ, 650, 604
\bibitem[]{} Riechers, D. A., Walter, F., Carilli, C. L., \& Bertoldi, F. 2007, ApJ, 671, L13
\bibitem[]{} Riechers, D. A., Walter, F., Carilli, C. L., Bertoldi, F., \& Momjian, E. 2008a, ApJ, 686, L9
\bibitem[]{} Riechers, D. A., Walter, F., Brewer, B. J., Carilli, C. L., Lewis, G. F., Bertoldi, F., \& Cox, P. 2008b, ApJ, 686, 851
\bibitem[]{} Riechers, D. A., et al. 2009, ApJ, 703, 1338
\bibitem[]{} Robson, I., Priddey, R. S., Isaak, K. G., \& McMahon, R. G. 2004, MNRAS, 351, L29
\bibitem[]{} Ryan-Weber, E. V., Pettini, M., Madau, P., Zych,
B. J., 2009, MNRAS, 395, 1476
\bibitem[]{}Seaquist, E. R., Ivison, R. J., \& Hall, P. J. 1995, MNRAS, 276, 867
\bibitem[]{} Simpson, C. 2005, MNRAS, 360, 565
\bibitem[]{} Shields, G. A., Menezes, K. L., Massart, C. A., \& Vanden
Bout, P. ApJ, 641, 683
\bibitem[]{} Smail, I., Ivison, R. J., Blain, A. W., \& Kneib, J.-P. 2002, MNRAS, 331, 495
\bibitem[]{} Solomon, P. M., Downes, D., Radford, S. J. E., \&  Barrett, J. W. 1997, ApJ, 478, 144
\bibitem[]{} Solomon, P. M., \& Vanden Bout, P. A. 2005, ARA\&A, 43, 677, {\bf SV05}
\bibitem[]{} Spergel, D. N. et al. 2007, ApJS, 170, 377
\bibitem[]{} Tacconi L. J. et al., 2006, ApJ, 640, 228
\bibitem[]{} Tremaine, S. et al. 2002, ApJ, 574, 740
\bibitem[]{} Treister, E., Krolik, J. H., \& Dullemond, C. 2008, ApJ, 679, 140
\bibitem[]{} Wagg, J., Hughes, D. H., Aretxaga, I., Chapin, E. L., Dunlop, J. S., Gazta$\rm \tilde{n}$aga, E., \& Devlin, M. 2007, MNRAS, 375, 745
\bibitem[]{} Wagg, J , Wang, R., Carilli, C., Walter, F., Maddalena, R. \& Pisano D.J. 2008, GBT memo \#256,
\bibitem[]{} Walter, F. et al. 2003, Nature, 424, 406
\bibitem[]{} Walter, F., Carilli, C. L., Bertoldi, F., Menten, K., Cox,
P., Lo, K. Y., Fan, X., \& Strauss, M. A. 2004, ApJ, 615, L17
\bibitem[]{} Walter, F., Riechers, D., Cox, P., Neri, R., Carilli, C.,
Bertoldi, F., Wei$\ss$, A., \& Maiolino, R. 2009, Nature, 457, 699
\bibitem[]{} Wang, R. et al. 2007, AJ, 134,
617
\bibitem[]{} Wang, R. et al. 2008a, AJ, 135, 1201
\bibitem[]{} Wang, R. et al. 2008b, ApJ, 687, 848
\bibitem[]{} Wei$\ss$, A., Downes, D., Walter, F., \& Henkel, C. 2005a, A\&A, 440, L45
\bibitem[]{} Wei$\ss$, A., Walter, F., \& Scoville, N. Z. 2005b, A\&A, 438, 533
\bibitem[]{} Wei$\ss$, A., Downes, D., Neri, R. et al. 2007, A\&A, 467, 955
\bibitem[]{} Willott, C. J., McLure, R. J. \& Jarvis, M. J. 2003, ApJ, 587, L15
\bibitem[]{} Willott, C. J. et al. 2007, AJ, 134, 2435
\bibitem[]{} Willott, C. J. et al. 2009a, AJ, 137, 3541
\bibitem[]{} Willott, C. J. et al. 2009b, AJ, in press., astro-ph/0912.0281
\bibitem[]{} Wu, X.-B. 2007, ApJ, 657, 177
\end{thebibliography}
\end{document}